\documentclass[12pt,a4paper,final]{iopart}
\usepackage{color,iopams,amssymb,array}
\usepackage{graphics,epstopdf}
\usepackage{hyperref}
\hypersetup{
    breaklinks=true
    bookmarks=true,         % show bookmarks bar?
    unicode=false,          % non-Latin characters in Acrobat's bookmarks
    pdftoolbar=true,        % show Acrobat's toolbar?
    pdfmenubar=true,        % show Acrobat's menu?
    pdffitwindow=false,     % window fit to page when opened
    pdfstartview={FitH},    % fits the width of the page to the window
    pdftitle={Statistics of the work distribution for a quenched Fermi gas},    % title
    pdfauthor={A. Sindona,J. Goold, N. Lo Gullo, F. Plastina},     % author
    pdfsubject={Work distribution and sudden quenching},   % subject of the document
    pdfcreator={LaTeX},   % creator of the document
    pdfproducer={PP}, % producer of the document
    pdfkeywords={key1} {key2} {key3}, % list of keywords
    pdfnewwindow=true,      % links in new window
    colorlinks=true,        % false: boxed links; true: colored links
    linkcolor=blue,          % color of internal links
    urlcolor=blue,        % color of links to bibliography
    filecolor=red,      % color of file links
    citecolor=blue     % color of external links
}
\usepackage{soul}

\newcommand{\text}[1]{\mathrm{#1}}

\newcolumntype{P}{>{\raggedright\arraybackslash} m{.175\linewidth}}

\newcommand{\w}{\omega}
\newcommand{\wo}{\omega_{0}}
\newcommand{\uo}{\omega_{0}^{\prime}}
\newcommand{\rp}{r^{\prime}}
\newcommand{\en}{\varepsilon_{n}}
\newcommand{\er}{\varepsilon_{2r}}
\newcommand{\hw}{\hbar \omega }
\newcommand{\tp}{t^{\prime}}
\newcommand{\ts}{t^{\prime{\hskip-0.5pt}\prime}}
\newcommand{\msc}[1]{\hbox{\uppercase{\tiny #1}}}
\newcommand{\eF}{\varepsilon_{\msc{f}}}
\newcommand{\rF}{r_{\msc{f}}}
\newcommand{\ee}{\mathrm{e}}
\newcommand{\dd}{\mathrm{d}}
\newcommand{\spind}{(2s+1)}
\newcommand{\spinwf}{\psi_{\xi}}
\newcommand{\wP}{\eta}
\newcommand{\kr}{\tilde{r}}
\newcommand{\ek}{\varepsilon_{2\kr}}

\newcommand{\fourIdx}[5]{
\setbox1=\hbox{\ensuremath{^{#1}}}
\setbox2=\hbox{\ensuremath{_{#2}}}
\setbox5=\hbox{\ensuremath{#5}}
\hspace{\ifnum\wd1>\wd2\wd1\else\wd2\fi}
\ensuremath{\copy5^{\hspace{-\wd1}\hspace{-\wd5}#1\hspace{\wd5}#3}_{\hspace{-\wd2}\hspace{-\wd5}#2\hspace{\wd5}#4}}
}
\newcommand{\HypF}{\fourIdx{ }{2}{ }{1}{F}}
\newcommand{\HypFR}{\fourIdx{ }{2}{ }{1}{\tilde{F}}}

\newcommand{\Exclude}[1]{}

\newcommand{\PwT}{P_{\wP_{0} \rightarrow \wP_{\tau}}(W)}
\newcommand{\PwQ}{P_{\wP_{-}\rightarrow \wP _{+}}(W)}

\begin{document}

\title{Statistics of the work distribution for a quenched Fermi gas}

\author{A.~Sindona$^{1,2}$}
\address{$^1$Dipartimento di Fisica, Universit\`a della Calabria, 87036 Arcavacata di Rende (CS), Italy\\}
\address{$^2$INFN sezione LNF-Gruppo collegato di Cosenza, Italy\\}
\ead{sindona@fis.unical.it}

\author{J.~Goold$^{3}$}
\address{$^3$The Abdus Salam International Centre for Theoretical Physics, 34014, Trieste, Italy\\}
\ead{jgoold@ictp.it}

\author{N.~Lo~Gullo$^{4,5}$}
\address{$^{4}$Quantum Systems Unit, Okinawa Institute of Science and Technology and Graduate University, Okinawa, Japan\\}
\address{$^{5}$Dipartimento di Fisica e Astronomia ``Galileo Galilei'', Universit\`a degli studi di Padova, Italy\\}
\ead{logullo@pd.infn.it}

\author{F.~Plastina$^{1,2}$}
\address{$^1$Dipartimento di Fisica, Universit\`a della Calabria, 87036 Arcavacata di Rende (CS), Italy\\}
\address{$^2$INFN sezione LNF-Gruppo collegato di Cosenza, Italy\\}
\ead{francesco.plastina@fis.unical.it}

\begin{abstract}
The local quench of a Fermi gas, giving rise to the Fermi edge
singularity and the Anderson orthogonality catastrophe, is a rare
example of an analytically tractable out of equilibrium problem in
condensed matter. It describes the universal physics which occurs
when a localized scattering potential is suddenly introduced in a
Fermi sea leading to a brutal disturbance of the quantum state. It
has recently been proposed that the effect could be efficiently
simulated in a controlled manner using the tunability of
ultra-cold atoms. In this work, we analyze the quench problem in a
gas of trapped ultra-cold fermions from a thermodynamic
perspective using the full statistics of the so called {\it work}
distribution. The statistics of work are shown to provide an
accurate insight into the fundamental physics of the process.
\end{abstract}

 \vspace{2pc} \noindent{\it Keywords}:
work distribution, non-equilibrium thermodynamics, quantum quench,
orthogonality catastrophe. \submitto{\NJP}

\section{Introduction}

In the past decade ultra-cold quantum gases have emerged as ideal
candidates for clean and controllable simulation of condensed
matter physics~\cite{Bloch:08}. Ultra-cold quantum gases are now
created in a variety of configurations in laboratories worldwide.
In particular, both Bosonic and Fermionic atoms can be trapped and
manipulated on optical lattice potentials~\cite{Lewenstein:07}.
The lack of thermal phonons coupled with the tunability of the
interactions by means of Feschbach resonances~\cite{Courteille:98}
has allowed for the detailed study of a multitude of phase
diagrams, the most celebrated example is perhaps the Bose-Hubbard
model~\cite{Jaksch:98,Greiner:02,Greiner2:02}.
\\
Equilibrium properties aside, over the past number of years there
has been a surge in interest in the out of equilibrium behavior of
closed quantum systems following a quench of a Hamiltonian
parameter. Fundamentally, this is due to a series of spectacular
experiments in ultra-cold atoms whereby the high degree of
isolation and long coherence times permits the study of dynamics
over long timescales~\cite{Greiner:02,Kinoshita2006}. These
experiments have raised a number of important theoretical issues
such as the relationship between thermalisation and integrability
and the universality of defect generation following evolution
across a critical point~\cite{silva}.
\\
Given the controllability of ultra-cold atomic systems and the
current interest in quench dynamics, it is a natural question to
ask if there are any out of equilibrium condensed matter physics
problems which could be simulated. Unfortunately, due to their
intrinsic complexity, there are very few examples of exactly
solvable problems in the out-of-equilibrium domain. A worthy
exception is the phenomenon of orthogonality
catastrophe~\cite{Anderson:67,AndersonYuval,mahan:00} and the
Fermi-edge singularity. This problem was first pointed out by
P.~W.~Anderson over 40 years ago when he showed that the overlap
of two many-body wave-functions, which describe deformed and
undeformed Fermi seas, vanishes in the thermodynamic
limit~\cite{Anderson:67}. The corresponding `quench' problem, was
investigated a few years later with the prediction of a universal
absorption-edge singularity in the X-ray spectrum of metals, the
`Fermi-edge' singularity~\cite{Nozieres}.
\\
The universal physics of the Anderson orthogonality catastrophe
and the Fermi-edge singularity was recently explored by Goold
\textit{et al.} in the context of ultra-cold quantum
gases~\cite{goold:11}. In this work it was suggested that the
physics maybe simulated in a controlled fashion by the appropriate
embedding of a single probe qubit. The approach was further
formalized in~\cite{sindona:12} where the authors solved the
dynamical problem in the inhomogeneous system by means of a linked
cluster expansion. This qubit probe approach was further suggested
as a mechanism to probe the physics of the orthogonality
catastrophe in~\cite{Knap:12,luttinger}.
\\
Interestingly enough, a connection was made by Heyl and
Kehrein~\cite{heyl} between the absorption and emission spectrum
in the original X-ray experiments and the so called quantum work
distribution and corresponding fluctuation relations in classical
and quantum statistical mechanics~\cite{jrev,mrev}. Treating a
quench problem in manybody physics as a thermodynamic
transformation and analyzing the statistics of work done has
recently shown to be a useful approach to understand the intrinsic
out of equilibrium dynamics in manybody
systems~\cite{Silva,dorner,Silva2,Sotiriadis,latent}. The approach
is based on the study of the moments of a quantity known as the
quantum work distribution~\cite{Tasaki,lutz} which have been found
to encode both thermodynamic and universal features of the model
in question.
\\
In this work this relationship will be explored in detail in the
context of a locally quenched trapped Fermi gas. In particular, in
section~\ref{sec100} the general problem of a system Hamiltonian
depending on a work-parameter will be introduced, adopting the
description based on the grand canonical ensemble \cite{gici}. The
formalism provided by the work distribution and its characteristic
function will be discussed, and the concept of irreversible work
will be explored. In Section~\ref{sec200} the focus will be moved
to  a Fermi gas in equilibrium with a harmonic trap, being
suddenly perturbed by a spatially structure-less perturbation. In
section~\ref{LCE} the vacuum persistence amplitude and the linked
cluster expansion will be used to reduce the calculation of the
characteristic function of work to the sum of connected Feynman
diagrams. The relation of these diagrams to the characteristic
function of work will be covered and an analytic approximation
holding at low temperature will be presented. In
section~\ref{sec500} the first three cumulants of the work
distribution will be computed and their link with thermodynamics
will be discussed. Finally, in section~\ref{sec600} the
irreversible work will be calculated using a perturbative and a
numerical approaches, while Section \ref{concluse} will provide
some further comments and conclusions.

\section{Non-equilibrium quantum thermodynamics}

\label{sec100}

Consider a system evolving according to the Hamiltonian
$\hat{H}(\wP)$, which depends on some  externally tunable
parameter $\wP$. The system is brought into weak contact with a
heat reservoir, at inverse thermal energy  $\beta$, and allowed to
equilibrate for each chosen value of $\wP$, before the
system-reservoir coupling is turned off. To each $\wP$ there
corresponds a well defined Gibbs state
\begin{equation}
\hat{\rho}(\wP)= \frac{\ee^{-\beta [\hat{H}(\wP)-\mu \hat{N}]}}{Z(\wP)},
\quad \text{with} \quad
Z(\wP)=\tr\left\{\ee^{-\beta [\hat{H}(\wP)-\mu \hat{N}]}\right\}
\label{rhoG/Zp}
\end{equation}
being the grand-canonical partition function and $\hat{N}$ the particle number operator.

Suppose that $\wP$ is some explicitly time-dependent degree of
freedom, which interacts directly with the system and not with the
reservoir, i.e., a \textit{work parameter} $\wP=\wP_{t}$. Then,
for an initial value $\wP_{0}$, at the time $t=0$, the state of
the system is $\hat{\rho}(\wP_{0})$. The initial Hamiltonian and
particle number operators have, respectively, the spectral
decompositions
\[
\hat{H}(\wP_{0})=\sum_{n}E_{n}(\wP_{0})|{n}\rangle \langle {n}|
{\quad }\text{and}{\quad }
\hat{N}=\sum_{n}N_{n}|{n}\rangle \langle {n}|\text{,}
\]
where $|{n}\rangle $ is the $n^{\text{th}}$ simultaneous
eigenstate of $\hat{H}(\wP_{0})$ and $\hat{N}$, with eigenvalues
$E_{n}(\wP_{0})$ and $N_{n}$. Next, some `work' is performed on
the system taking the work parameter from $\wP_{0}$ to a final
value $\wP_{\tau }$,  at a later time $t=\tau $. The final
Hamiltonian, connected by the protocol $\wP_{0}\rightarrow
\wP_{\tau }$, has the spectral decompositions
\[
\hat{H}(\wP_{\tau })=\sum_{m}E_{m}^{\prime}(\wP_{\tau })|{m}\rangle
\langle {m}| \text{,}
\]
where $|{m}\rangle $ is the $m^{\text{th}}$ simultaneous
eigenstate of $\hat{H}(\wP_{\tau })$ and $\hat{N}$, with
eigenvalues $E_{m}^{\prime}(\wP_{\tau })$ and $N_{m}$.

The definition of work in this scenario requires two projective
measurements: the first projects onto the eigenbasis of the
initial Hamiltonian $\hat{H}(\wP_{0})$, with the system in thermal
equilibrium. The system then evolves under the unitary dynamics
$U(\tau ,0)$, generated by the protocol $\wP_{0}\rightarrow
\wP_{\tau }$, before the second measurement projects onto the
eigenbasis of the final Hamiltonian $\hat{H}(\wP_{\tau })$. The
probability of obtaining $E_{n}(\wP_{0})$ for the first
measurement outcome followed by $E_{m}^{\prime}(\wP_{\tau })$ for the
second is then
\begin{equation}
p_{n}^{0}\,p_{m|n}^{\tau } =
\frac{1}{Z(\wP_{0})}
\ee^{-\beta\,[E_{n}(\wP_{0})-\mu N_{n}]}
|{\langle n|}U(\tau ,0){|m\rangle }|^{2}.
\label{pnpnm}
\end{equation}
The work distribution is defined as~\cite{Tasaki,lutz}
\begin{equation}
\PwT=\sum_{n,m} p_{n}^{0}\,p_{m|n}^{\tau }\,\delta [
W-E_{m}^{\prime}(\wP_{\tau})+E_{n}(\wP_{0})], \label{eq:qworkdist}
\end{equation}
and the average work $\langle W\rangle$ done on the system is given by the first moment of $\PwT$.
It is useful to introduce the characteristic function of work~\cite{lutz} as
\begin{eqnarray}
\chi (t,\tau ) &=&\langle e^{\frac{it}{\hbar }W}\rangle
=\int \dd{W}\,\ee^{\frac{it}{\hbar }W}\PwT  \label{eq:loschmidt} \\
&=&\left\langle
U^{\dag }(\tau ,0)\ee^{\frac{it}{\hbar }\hat{H}(\wP_{\tau})}
U(\tau ,0)\ee^{-\frac{it}{\hbar }\hat{H}(\wP_{0})}\right\rangle ,
\nonumber
\end{eqnarray}%
with $\langle {\cdots }\rangle =\tr\lbrack {\cdots }\hat{\rho}(\wP_{0})]$ denoting the thermal equilibrium average over the initial state.
In terms of $\chi (t,\tau )$ the average work is expressed as $\langle W\rangle =-i\hbar\,d\chi (t,\tau )/dt|_{t=0}$.

Microscopically, the second law of thermodynamics is revised to
the form \[\langle W\rangle \geq \Delta\Omega
=-\ln[{Z(\wP_{\tau })/Z(\wP_{0})}]/\beta,\] with $\Omega(\wP) = - \ln[Z(\wP)]/\beta$ being the thermodynamic grand potential, so as to encompass the
explicit statistical nature of work. The deficit between average
work and the variation in the grand potential can be accounted for
by the introduction of the irreversible work contribution $\langle
W_{\msc{irr}}\rangle > 0$, being such that $\langle W\rangle
=\Delta {\Omega }+\langle W_{\msc{irr}}\rangle$. When combined
with the first law of thermodynamics, this relation can be
rewritten as
\[
\Delta S=\Delta S_{\msc{rev}}+\Delta S_{\msc{irr}},
\]
where $\Delta S$ is the change in entropy of the system and
$\Delta S_{\msc{irr}}=\beta \langle W_{\msc{irr}}\rangle$~($\Delta
S_{\msc{rev}}=\beta\langle \Delta Q\rangle$) is the
irreversible~(reversible) entropy change. We note that for a
closed quantum system, the heat transfer into the system is zero
and the sole contribution to the entropy change is the
irreversible entropy. To these concepts and their link to the
irreversibility of a sudden transformation, we shall come back in
Sec.~\ref{sec600}, after we have evaluated the characteristic
function of the work distribution and its first moments for the
specific problem at hand.

\section{The Locally Perturbed Fermi Gas}
\label{sec200}

\subsection{Model Hamiltonian}

Consider a gas of non-interacting cold fermions, confined by a one-dimensional trapping harmonic potential.
The trap has a characteristic length $x_{0}$ and a resonance frequency $\w$.
Then, each particle of the gas, in its unperturbed state, is described by the harmonic oscillator Hamiltonian
\begin{equation}
H_{0}(x)=\frac{1}{2}
\left(
-\frac{x_{0}^{2}\partial ^{2}}{\partial x^{2}}
+\frac{x^{2}}{x_{0}^{2}}
\right).
\label{H0}
\end{equation}
The unperturbed system Hamiltonian
\begin{equation}
\hat{H}_{0}=\sum_{\xi }\int {\dd}x\,
\hat{\Psi}_{\xi }^{\dag}(x)\,H_{0}(x)\,\hat{\Psi}_{\xi }(x),
\label{Hhat0}
\end{equation}
is spanned by the fermion field in real space $\hat{\Psi}_{\xi }(x)$, with $\xi$ denoting the spin degrees of freedom, i.e., $\xi$ ranging over $\spind$ values.

Let us now consider doing \textit{work} on the Fermi gas by the turning on of an external potential.
Let us further assume that the switch is done in a sudden way.
In this paper we will take this external perturbation to be a localized potential at the centre of the trap which we can model by a Dirac $\delta $ function with strength $V_{0}$ such that
$V(x,t)=\theta (t)\pi V_{0}x_{0}\delta (x)$,
where $\theta (t)$ is the Heaviside step function.
Then, the total Hamiltonian for the gas after the switching on of the potential is $\hat{H}=\hat{H}_{0}+\hat{V}$, where
\begin{equation}
\hat{V}=\pi V_{0}x_{0}\,\sum_{\xi }
\hat{\Psi}_{\xi }^{\dag}(0)\,\hat{\Psi}_{\xi }(0).
\label{Vhat}
\end{equation}

\subsection{The characteristic function of work and the vacuum persistence
amplitude}
Consider the work parameter
$\wP_{t}=\theta (t)\pi V_{0}x_{0}$
and the protocol
$\wP_{-}\rightarrow \wP_{+}$,
from $t\rightarrow 0_{-}$~(with $\wP_{-}=0$) to $t\geq 0_{+}$~(with $\wP_{+}=\pi V_{0}x_{0}$).
Then, the initial state of the system $\hat{\rho}(\wP_{-})$ is given by Eq.~(\ref{rhoG/Zp}) with the initial Hamiltonian $\hat{H}(\wP_{-})=\hat{H}_{0}$,
introduced in Eq.~(\ref{Hhat0}), and the particle number operator
$\hat{N}=\sum_{\xi }\hat{\Psi}_{\xi }^{\dag }(0)\,\hat{\Psi}_{\xi }(0)$.
The partition function in the initial state will be denoted $Z$.
The final Hamiltonian $\hat{H}(\wP_{+})=\hat{H}_{0}+\hat{V}$ includes the perturbation operator introduced in Eq.~(\ref{Vhat}).

In this particular case a sudden quench occurs, and the characteristic function of work~(\ref{eq:loschmidt}) reduces to~\cite{lutz}
\begin{equation}
\chi _{\beta }(t)= \left\langle
e^{\frac{it}{\hbar }\hat{H}(\wP_{+})}\,
\ee^{-\frac{it}{\hbar }\hat{H}(\wP_{-})} \right\rangle.
\label{Eq:sudden}
\end{equation}
In the dynamical response theory of many particle system, a key quantity to calculate is the so-called \textit{vacuum persistence amplitude}~\cite{colemann}
\begin{equation}
\nu _{\beta }(t>0)=\left\langle {e^{\frac{i}{\hbar }\hat{H}_{0}t}{\,} e^{-%
\frac{i}{\hbar }(\hat{H}_{0}+\hat{V})t}} \right\rangle,  \label{Eq:VacAmp}
\end{equation}
i.e., the probability amplitude that the gas will retrieve its equilibrium state at time $t$, after the switching on of the perturbation.
It is quite obvious that a simple relationship exists between the characteristic function~(\ref{Eq:sudden}) and the vacuum persistence amplitude~(\ref{Eq:VacAmp}):
\begin{equation}
\chi _{\beta }(t)=\nu _{\beta }^{\ast }(t).  \label{chinu}
\end{equation}
Exploiting this relationship we have the full access to the statistics of work done via calculation of the vacuum persistence amplitude.

Initially the fermions lie in their equilibrium configuration, set by $\hat{H}_{0}$, until a sudden perturbation $\hat{V}(t)=\hat{V}\theta (t)$ is felt by the gas.
The initial equilibrium depends on the harmonic oscillator Hamiltonian given in Eq.~(\ref{H0}) whose unperturbed eigenfunctions are
\[
\psi _{n}(x)=\frac{x_{0}^{-1/2}\pi ^{-1/4}}{2^{n/2}n!^{1/2}}{\,}
H_{n}\!\left( \frac{x}{x_{0}}\right) {\,}e^{-\frac{x^{2}}{2x_{0}^{2}}},
\]
with $H_{n}(x/x_{0})$ being the Hermite polynomials of order $n$.
By the parity of these wave~functions, the external potential
induces excitations which connect only unperturbed one-fermion
states labeled by even numbers $n=2r$, with $r=0,1,\cdots ,\infty
$. Then, the matrix elements of the external potential in the
unperturbed oscillator basis read
\begin{equation}
V_{r\rp}={\pi }V_{0}x_{0}\,\psi _{2r}^{\ast }(0)\,
\psi _{2\rp}(0)=V_{0}\,(-1)^{r+\rp}\,\gamma _{r}^{1/2}\,\gamma _{\rp}^{1/2},
\label{Vrrp}
\end{equation}
where we have introduced the Euler gamma function ratio $\gamma _{r}=\Gamma (r+1/2)/\Gamma (r+1)$.
We express the fermion field in terms of the annihilation operator $\hat{c}_{n\xi }$ for the (unperturbed) $n$-th single
particle state of energy $\en={\hw}(n+1/2)$ and spin $\xi$.
Then, using
\[
\hat{\Psi}_{\xi }(x)=\sum_{n}\psi _{n}(x){\spinwf}\hat{c}_{n\xi },
\]
with ${\spinwf}$ being the spinor wave~function, the unperturbed Hamiltonian results in
\begin{equation*}
\hat{H}_{0}=\sum_{n,\xi }\en\hat{c}_{n\xi }^{\dag }\hat{c}_{n\xi}
=\sum_{r,\xi }\er\hat{c}_{2r\,\xi }^{\dag }\hat{c}_{2r\,\xi }
+\sum_{r,\xi}\varepsilon_{2r+1}\hat{c}_{2r+1\,\xi }^{\dag }\hat{c}_{2r+1\,\xi},
\end{equation*}
the particle number operator reads
$
\hat{N}=\sum_{n,\xi }\hat{c}_{n\xi }^{\dag }\hat{c}_{n\xi}\text{,}
$
while the effect of the perturbation on the gas is represented by
\begin{equation*}
\hat{V}=\sum_{r,\rp,\xi }V_{r\rp}\hat{c}_{2r\xi }^{\dag }\hat{c}_{2\rp\xi }.
%\label{VhatD}
\end{equation*}

\section{Linked Cluster expansion}
\label{LCE}

A typical approach in many-body physics to find a manageable
expression for the vacuum persistence amplitude~(\ref{Eq:VacAmp})
is to turn to the interaction picture, where the impurity
potential reads
\[
\tilde{V}(t)=
{\ee}^{\frac{i\,t}{\hbar } \hat{H}_{0}}
\,\hat{V}\,
{\ee}^{-\frac{i\,t}{\hbar} \hat{H}_{0}}.
\]
In this representation, we may use the identity~(\ref{chinu}) to express the characteristic function of work as
\begin{equation}
\chi _{\beta }(t)=
\left\langle
T \ee^{\frac{i}{\hbar }
\int_{0}^{t}\dd\tp\,\tilde{V}(\tp)}
\right\rangle \text{.}  \label{IP}
\end{equation}
Then, we may perform a linked cluster expansion and reduce $\chi _{\beta }(t)$ to an exponential sum of connected Feynmann diagrams:
\begin{equation}
\begin{array}{P P}
$\chi_{\beta}(t)=\ee^{\Lambda^{\ast}_{\beta }(t)}$,
&
\scalebox{0.91}{\includegraphics{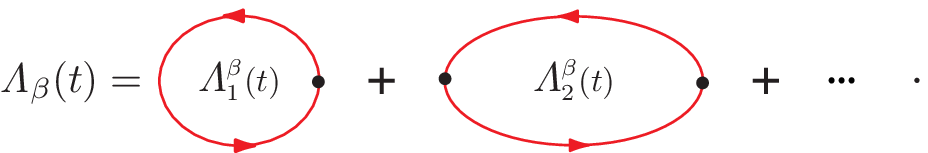}}
\end{array}
\label{nubetF}
\end{equation}

\begin{figure}[tbh]
%\raggedleft
\hskip 6pc \scalebox{0.99}{\includegraphics{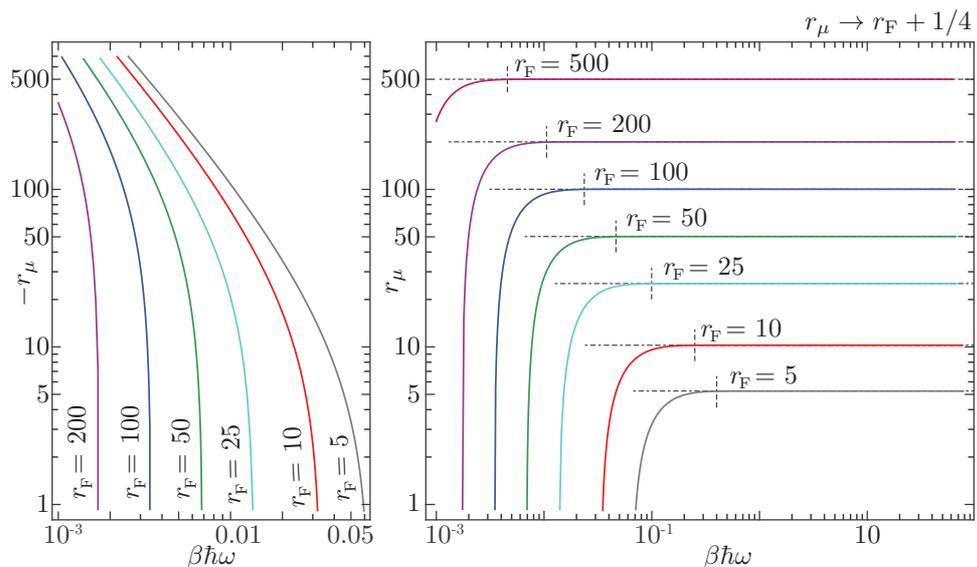}} \vskip
-12pt \caption{Behaviour of the chemical potential index $r_{\mu
}$ vs $\beta\hw$ for spin $1/2$ gases with number of particles
$\langle\hat{N}\rangle =2(2\rF+1)$ in the range of $22$ to $2002$.
$r_{\mu}$ and hence $\mu =\hw(2r_{\mu }+1/2)$ are close to their
maximum values, i.e., $r_{\mu}\rightarrow \rF+1/4$ and $\mu
\rightarrow \hw(2\rF+1)$, for $\beta \hw\gtrsim 0.004-0.4$
depending on $\rF$. On the other hand, for $\beta
\hw=10^{-3}-10^{-1}$ they show an abrupt decrease to large
negative values. } \label{figOne}
\end{figure}

\noindent These loops contain separable products of lines
\begin{equation}
i\hbar G_{r}^{\beta }(t)=
\ee^{-i{\er}t/\hbar }[
\theta(t)\,f_{r}^{-}-\theta (-t)\,f_{r}^{+}
],  \label{prop}
\end{equation}
connected to vertices~($V_{rr^{\prime }}$), with level occupation numbers given by the Fermi-Dirac distributions
\begin{equation}
f_{r}^{+}=\langle \hat{c}_{2r\,\xi }^{\dag }\hat{c}_{2r\,\xi}\rangle =\frac{1}{1+\ee^{\beta (\er-\mu )}},
\quad
f_{r}^{-}=\langle \hat{c}_{2r\,\xi }\hat{c}_{2r\,\xi }^{\dag}\rangle =\frac{1}{1+\ee^{-\beta (\er-\mu )}}.
\label{FDD}
\end{equation}
As is standard in discrete-level systems, and intrinsic
semiconductors, we let the chemical potential $\mu$ lie in the
middle between the Fermi energy $\eF$ and the lowest unoccupied
one-fermion state at the absolute zero~($\beta \rightarrow \infty
$). Without loss of generality, we may assume the Fermi level
number $n_{\msc{f}}$ to be even, label it by $2\rF$~($\rF$ being a
positive integer) so that $\eF=\hw(2\rF+1/2)$. Then, we can
parametrize $\mu $ as $\mu =\hw(2r_{\mu }+1/2)$, i.e., $f_{r}^{\pm
}=[1+\ee^{\pm 2\beta \hw(r-r_{\mu })}]^{-1}$, and compute
$r_{\mu}$ for finite $\beta $ by constraining the average particle
number to be:
\begin{equation}
\langle \hat{N}\rangle
=\sum_{r,\xi} (f_{r}^{+}+f_{r+1/2}^{+})
=\spind(2\rF+1)\text{.}  \label{NvsSpinAndRF}
\end{equation}%
From the plots of Fig.~\ref{figOne}, we observe that $\mu$ takes its maximum value
\begin{equation}
\mu_{\infty }=
\frac{\hw\langle \hat{N}\rangle }{\spind}=\hw(2\rF+1),
\qquad
r_{\mu} \to \rF+1/4,  \label{muinf}
\end{equation}
at $\beta \hw\rightarrow \infty $. Then, it decreases with
decreasing $\beta\hw$, and it becomes largely negative for very
low $\beta \hw$ where the classical limit applies.

\noindent We now apply Wick's theorem and focus on the lowest
order loops in~(\ref{nubetF}), i.e., $\Lambda _{1}^{\beta }(t)$
and $\Lambda _{2}^{\beta}(t)$. In terms of the auxiliary functions
\begin{equation}
\lambda_{\pm }^{\beta }(t)=
\sum_{r=0}^{\infty }\gamma _{r}\ee^{\pm 2 i r \w t}f_{r}^{\pm },
\label{lambdas}
\end{equation}
we express these contributions as
\begin{eqnarray}
\Lambda_{1}^{\beta }(t) &=&
-\frac{i\,t}{\hbar} \spind V_{0}\, \lambda_{+}(0),
\label{Lambda1}
\\
\Lambda _{2}^{\beta }(t) &=&
-\frac{\spind V_{0}^{2}}{\hbar ^{2}}
\int_{0}^{t}\dd\tp\,
\int_{0}^{\tp}\dd\ts\,
\lambda _{+}^{\beta }(\ts){\,}\lambda_{-}^{\beta }(\ts)
\text{.}
\label{Lambda2}
\end{eqnarray}
It is useful for the following to parameterize $V_{0}$ in terms of
the energy scale $\sqrt{2\hw\eF}$, which results from the
geometric mean between the spacing of even energy levels and the
Fermi energy. To this end, we introduce
\begin{equation}
\alpha =\frac{\spind V_{0}^{2}}{2\hbar \w\eF}  \label{alphaPP}
\end{equation}
as a sensible parameter, and to keep the interaction potential
small, we investigate the range $\alpha =0-1$.

It turns out that, in the free gas limit ($\omega \rightarrow 0$),
this interaction strength coefficient reduces to the so called
``critical parameter'' of the Mahan Nozi\`{e}res De Dominicis~(MND) theory of the edge singularity \cite{mahan:00,Nozieres},
which, in the X-ray absorption problem, also gives a measure of
the asymmetry of the absorption spectrum. Analogously, we will
find below, for our trapped gas case, that the parameter $\alpha$
determines the skewness of the work distribution.

\subsection{Connected diagrams and work distribution}
\label{Sec4.1}

The full derivation of Eqs.~(\ref{Lambda1}) and~(\ref{Lambda2}),
and the numerical calculations involved, can be found in a recent
paper by the authors~\cite{sindona:12}, and the reader interested
in specific details is directed there. In the present context, we
quote the main results and apply them to the determination of the
work distribution $\PwQ$ and its characteristic
function~(\ref{nubetF}).

\noindent The one-vertex loop is just the adiabatic response of the gas, being of the form
$\Lambda _{1}^{\beta }(t)=-{i} t E_{1}^{\beta }/\hbar $,
where
\begin{equation}
E_{1}^{\beta } = \spind V_{0} \lambda _{+}(0)
=\sqrt{2\spind\hw\eF\alpha}
\sum_{r=0}^{\infty }\gamma_{r}f_{r}^{+}
\label{eq:E1}
\end{equation}
represents the first-order correction to the equilibrium energy,
given by
\[
E_{0}^{\beta } = \langle \hat{H}_{0}\rangle =
\spind\sum_{r} (\er\,f_{r}^{+} + \varepsilon _{2r+1}\,f_{r+1/2}^{+})
\text{.}
\]
Eq.~(\ref{eq:E1}) brings a phase factor to $\chi _{\beta }(t)$, which corresponds to shifting the work distribution $\PwQ$ by $E_{\beta }^{1}$.

\noindent The two-vertex loop can be split into three parts with well defined trends and physical meaning, namely,
\begin{equation}
\Lambda _{2}^{\beta }(t) =
\Lambda _{2\msc{s}}^{\beta }(t) +
\Lambda _{2\msc{g}}^{\beta }(t) +
\Lambda _{2\msc{p}}^{\beta }(t).
\label{L2decomp}
\end{equation}
The first one is
$\Lambda _{2\msc{s}}^{\beta }(t)=-{i} tE_{2}^{\beta }/\hbar$,
where
\begin{equation}
E_{2}^{\beta } = \alpha\eF \sum\limits_{r{\neq }\rp} \varphi _{r\rp}f_{r}^{+}f_{\rp}^{-},
\quad
\varphi _{r\rp}=\frac{\gamma _{r}{\,}\gamma _{\rp}}{r-\rp}
\label{Delta2B}
\end{equation}
provides the second-order correction to $E_{0}^{\beta }$ and brings a further shift to $\PwQ$.
The second one
$\Lambda _{2\msc{g}}^{\beta }(t)=-\delta _{\beta }^{2} \w  ^{2}t^{2}/2$
includes the coefficient
\begin{equation}
\delta _{\beta } = \sqrt{2\alpha g_{\beta }},\quad
g_{\beta }=\frac{\eF}{\hw} \sum_{r}\gamma _{r}^{2}\,f_{r}^{+}\,f_{r}^{-},
\label{eq:gbeta}
\end{equation}
which produces a Gaussian damping in $\chi _{\beta }(t)$ and a Gaussian broadening in $\PwQ$.
The third one
\begin{equation}
\Lambda _{2\msc{p}}^{\beta }(t)= -\frac{\alpha \eF}{2\hw}
\sum\limits_{r{\neq }\rp}
\psi _{r\rp}(t)\,f_{r}^{+}\,f_{\rp}^{-},
\quad
\psi _{r\rp}(t)=\varphi _{r\rp}{\,}\frac{1-\ee^{2{i} (r-\rp)t \w }}{r-\rp}
\label{Lambda2E}
\end{equation}
accounts for the shake-up of the gas due to the sudden switching
of the impurity. It is a periodic function of time with frequency
$2 \w  $~(as a direct consequence of the harmonic form of trapping
potential), which provides the non trivial part of $\PwQ$.
\begin{figure}[tbh]
%\raggedleft
\hskip 6pc \scalebox{0.99}{\includegraphics{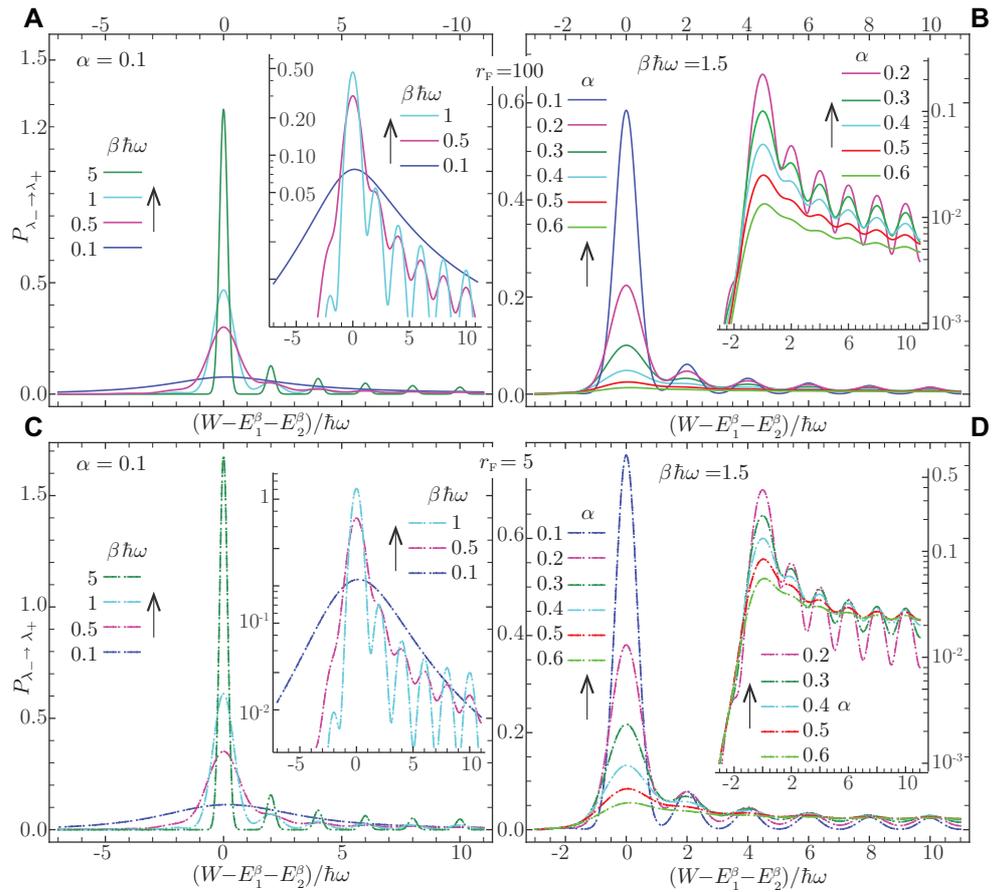}}
\vskip -12pt
\caption{Work distributions~(\ref{Pwork}) vs
$(W-E_{1}^{\beta }-E_{2}^{\beta })/\hw$
for different temperatures, such that $\beta\hw=0.1-5$,
particle numbers $\langle \hat{N}\rangle =22,202$, i.e., $\rF=5,100$,
and critical exponents $\alpha =0.1-0.6$.
The distributions shown in both linear and log scales were obtained by a fast Fourier transform algorithm on the numerical data from the `unshifted' characteristic function
$\chi ^{\prime }(t)=\ee^{-i(E_{1}^{\beta }+E_{2}^{\beta })t/\hbar }\chi (t)$~[see Eq.~(\ref{nust})].}
\label{figTtwo}
\end{figure}

The basic quantities of the problem, given in Eqs.~(\ref{eq:E1}),
(\ref{Delta2B}), (\ref{eq:gbeta}), and (\ref{Lambda2E}), contain
summations running over all one-fermion eigenstates of the trap,
weighted by the Fermi factors $f_{r}^{\pm }$, and enter the
characteristic function of work~(\ref{chinu}):
\begin{equation}
\chi _{\beta }(t) =
{\ee}^{\frac{it}{\hbar}(E_{1}^{\beta }+E_{2}^{\beta })}
{\ee}^{-\frac{\delta_{\beta }^{2}}{2}\w^{2}t^{2}}
{\ee}^{\Lambda _{2\msc{p}}^{\beta \ast }(t)}.
\label{nust}
\end{equation}
By Eq.~(\ref{eq:qworkdist}), the work distribution is given by the convolution product
\begin{equation}
\PwQ=\frac{
{\ee}^{-\frac{W^{2}}{2\,\delta_{\beta }^{2}}}}{\sqrt{2\pi }\delta _{\beta }}
\otimes
\int_{-\infty}^{\infty} \frac{\dd t}{2\pi \hbar}
{\ee}^{-\frac{it}{\hbar }(W-E_{1}^{\beta }-E_{2}^{\beta })}
{\ee}^{\Lambda _{2\msc{p}}^{\beta \ast }(t)}\text{.}
\label{Pwork}
\end{equation}
Numerical computations of $\PwQ$ are shown in Fig.~\ref{figTtwo},
where we recognize an asymmetric, broadened profile, signature of
the singular behavior of the Fermi gas. The monotonic structure
turns into a satellite structure of sub-peaks, separated by $2\hw$
and related to even-level transitions in the gas, as $\beta \hw$
gets above $\sim 0.5$.

These secondary peaks are a direct manifestation of the single
particle transitions occurring within the gas, with the fermions
jumping between even harmonic oscillator states (separated in
energy by even multiples of $2\hbar \omega$), in response to the
local perturbation. All of these transition constitute the so
called ``shake-up'' process, which is, thus, explicitly witnessed
by the work distribution.

At high temperatures, this feature disappears, being hidden by the
increased gaussian broadening of the main peak, which (in our
two-vertex approximation) is found at $W= E_1^{\beta} +
E_2^{\beta}$, and which describes the transition between the
equilibrium states induced by the switching of the scattering
center.

\subsection{Low thermal energy approximation}
\label{Sec4.2} Consider now a Fermi gas with a large number of
particles at sufficiently low temperatures, being such that we may
approximate the chemical potential by Eq.~(\ref{muinf}). In this
regime, we can expand $f_{r}^{\pm }$ in power series of $\ee^{\pm
\beta \hw(\er-\mu _{\infty })}$ and select the lowest order of
this series. As detailed in the appendix, we can work in the $\rF
\gg\ 1$-limit and find some manageable expressions for the
auxiliary functions $\lambda _{\pm }^{\beta }(t)$ introduced in
Eq.~(\ref{lambdas}). By Eq.~(\ref{Lambda1}), the initial value
$\lambda _{+}^{\beta }(0)$ determines the one-vertex loop and
hence the first order energy shifts~(\ref{eq:E1}). On the other
hand, the product $\lambda _{+}^{\beta }(t)\lambda _{-}^{\beta
}(t)$ enters a double time-ordered integral which
gives~(\ref{Lambda2}), together with the second-order energy
shifts~(\ref{Delta2B}), the Gaussian standard
deviation~(\ref{eq:gbeta}), and the shake up
sub-diagram~(\ref{Lambda2E}). This integral can be carried out
analytically, in the $\rF \gg\ 1$-limit, by adding a small shift
to the time domain on the imaginary axis. We, therefore, obtain
the following analytical approximation for the characteristic
function~(\ref{nust})
\begin{equation}
\chi _{\beta }(t) \approx
{\ee}^{\frac{{i}t}{\hbar }(E_{1}^{\infty}+E_{2}^{\infty })}
{\ee}^{-\frac{\delta _{\beta }^{2}}{2}\w^{2}t^{2}}
\left(
\frac{{\ee}^{2\tau_{0}\w}-1}
{{\ee}^{2\w\tau_{0}-2i\w t}-1}
\right)^{\alpha},
\label{nustinf}
\end{equation}
where
\begin{eqnarray}
E_{1}^{\infty } &=& 2\sqrt{\alpha \spind}\eF,
\qquad
E_{2}^{\infty }=\frac{-2\alpha \hw}{{1}-{\ee}^{-2\tau _{0}\w}},
\label{E12inf} \\
\delta _{\beta } &=&\sqrt{2\alpha g_{\beta }},\qquad g_{\beta}
=\sum_{m=1}^{M\rightarrow \infty }(-1)^{m+1}m\frac{{\ee}^{m\beta \hw/2}}{{\ee}^{m\beta \hw}-1}.
\label{gbetapp}
\end{eqnarray}
Here, $\tau_{0}$ is a regularization parameter, i.e., an imaginary
time shift, that we may interpret as the typical time interval
needed by the system to respond to the abrupt switching on of the
impurity potential at zero temperature. We see that the
function~(\ref{nustinf}) is undefined in the $\tau_0 \to 0$-limit,
which is a consequence of the large $\rF$ expansion used in the
thermal series for $\lambda _{\pm }^{\beta }(t)$. The exact
expressions provided by Eqs.~(\ref{Lambda1}) and~(\ref{Lambda2})
are well set and can be used to compute~(\ref{nust}) and
then~(\ref{Pwork}) numerically, without any divergence problem, as
we did in Fig.~\ref{figTtwo}.

We remark that the imaginary time regularization leading to
Eq.~(\ref{nustinf}) is required to provide an analytical support
to the theory. Furthermore, it has been observed
that~\cite{sindona:12}
\begin{equation}
e^{\Lambda _{2\msc{p}}^{\infty }(t)}=
\left( \frac{{\ee}^{2\tau _{0}\w}-1}{{\ee}^{2\w\tau _{0}+2\w {i}t}-1}\right)^{\alpha },
\label{LambXX}
\end{equation}
correctly tends to the Nozi\`{e}res and De~Dominicis core hole
propagator~\cite{Nozieres} when the harmonic trap frequency is
lowered to zero, keeping the number of particles in the gas
finite. In this limit, the regularization parameter $\tau_0$,
becomes exactly the one needed in the MND theory \cite{doniach}.
The mathematical details of the derivation of Eqs.~(\ref{E12inf}),
(\ref{gbetapp}), and~(\ref{LambXX}) are discussed  in~\ref{AppX1}.
\begin{figure}[tbh]
\hskip 6pc \scalebox{0.99}{\includegraphics{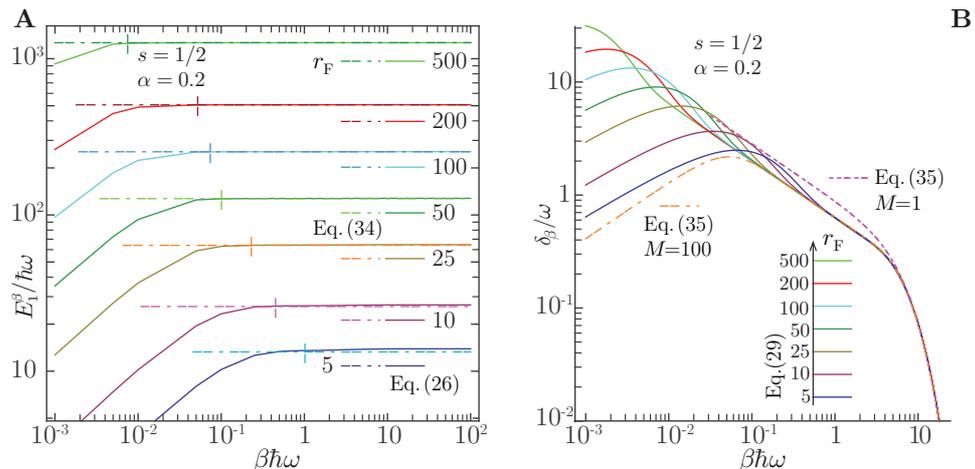}}
\vskip-12pt
\caption{(color on line)
First-order energy shift $E_{\beta }^{1}$ [Eq.~(\ref{eq:E1}), panel \textbf{A}]  and Gaussian damping coefficient~$\delta _{\beta }~$[Eq.~(\ref{eq:gbeta}), panel \textbf{B}] of a spin-$1/2$ gas into a harmonic trap interacting with a sudden, $\delta $-like potential.
The dimensionless quantities $E_{\beta }^{1}/\hw$ and $\delta _{\beta }/\w$ are plotted vs $\beta \hw$ and compared with the absolute zero approximations computed from Eq.~(\ref{E12inf}) and~(\ref{gbetapp}), respectively.
The critical exponent $\alpha $ is fixed to $\alpha =0.2$, while several particle numbers are considered in the range $\langle \hat{N}\rangle =22-1002$, i.e., $\rF=5-500$.}
\label{FigThree}
\end{figure}

\noindent The leading behavior of $\chi_{\beta }$
in~Eq.(\ref{nustinf}) vs temperature is provided by the gaussian
damping factor, whereas both the energy shifts and the periodic
part of the characteristic function are well approximated by their
absolute zero expressions in a wide range of temperatures,
corresponding to $\beta \hw\gtrsim 0.2$ for $\rF\gtrsim 10$. As
shown in Fig.~\ref{FigThree}\textbf{A}, the first-order shift
$E_{1}^{\infty }$, given by Eq.~(\ref{E12inf}), is the correct
large $\beta \hw$-limit for $E_{1}^{\beta }$, numerically
calculated from Eq.~(\ref{eq:E1}).
\begin{figure}[tbh]
\hskip6pc \scalebox{0.99}{\includegraphics{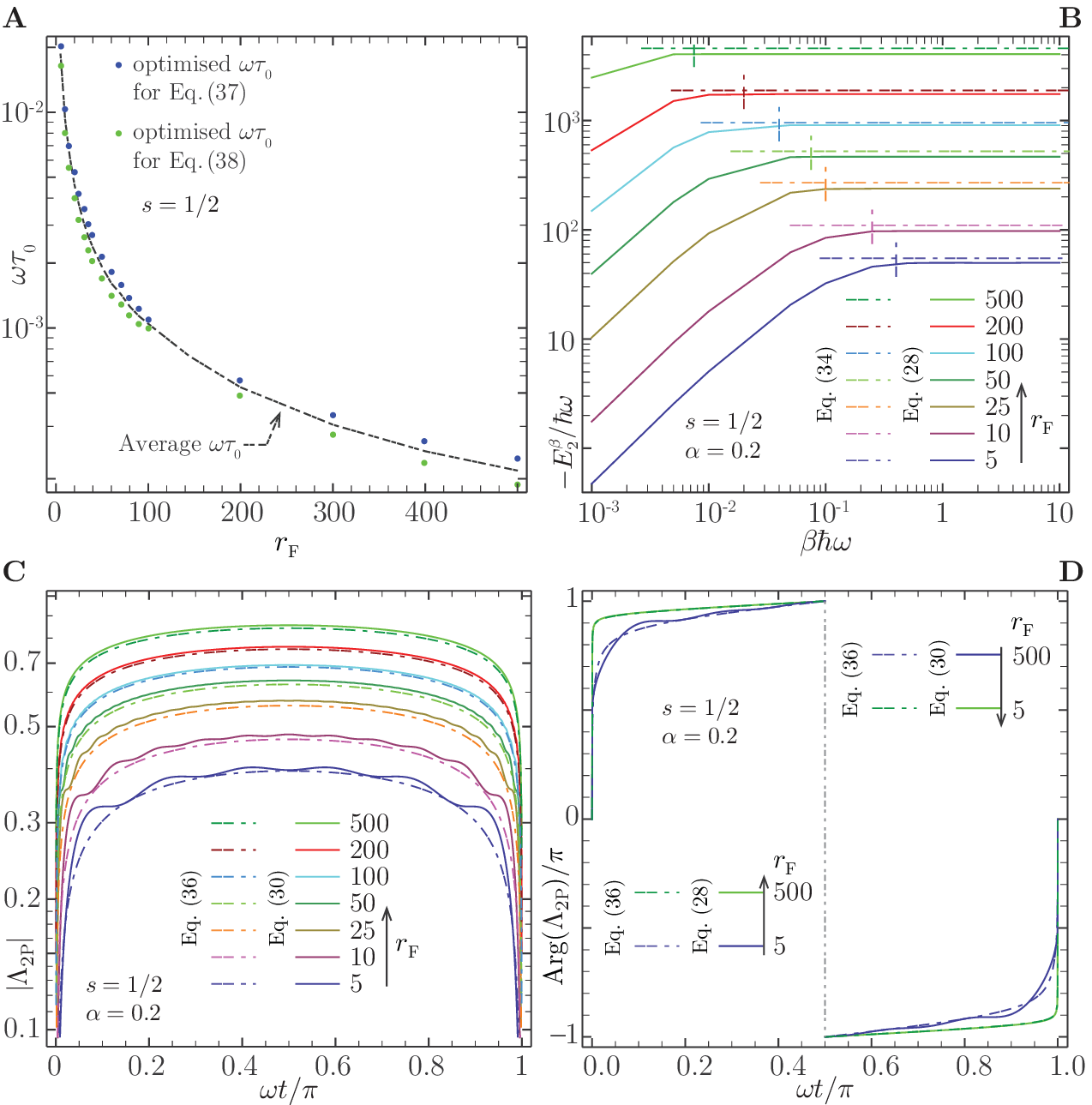}} \vskip-12pt
\caption{(color on line) (\textbf{A}) adjusted values of the
regularization parameter $\tau _{0}$ vs $\rF$~(Blue and Green
dots) that fulfill the conditions~(\ref{E2L2limA})
and~(\ref{E2L2limB}). The average between the two sets of values
is also reported. (\textbf{B}) Second-order energy shift $E_{\beta
}^{2}$ of a spin-$1/2 $ gas vs $\beta \hw$ and absolute zero
approximation computed from Eq.~(\ref{E12inf}), with the average
$\tau _{0}$ values of panel (\textbf{A}) for $\alpha =0.2$.
Modulus~(\textbf{C}) and phase~(\textbf{D}) of the periodic
sub-diagram $\Lambda _{2\msc{p}}^{\infty}$ vs $\w\,t$. Numerical
calculations from Eq.~(\ref{Lambda2E}) are compared with the
regularized expression given in Eq.~(\ref{LambXX}) where the
average $\tau _{0}$ values of panel~(\textbf{A}) are used. }
\label{Ffour}
\end{figure}

\noindent To compute the low temperature form of $\delta_{\beta
}$, we truncate the series in Eq.~(\ref{gbetapp}) and include the
first $M$ terms. The corresponding approximations, for $M=1$ and
$M=100$, are plotted in Fig.~\ref{FigThree}\textbf{B} together
with the numerical form of $\delta _{\beta}$ calculated from
Eq.~(\ref{eq:gbeta}). We see that already the $M=1$ curve
accurately reproduces the numerical data for $\beta \hw\gtrsim 5$.
The approximation with $M=100$ components works particularly well
in the extended range $\beta \hw\gtrsim 0.5$, where the numerical
$\delta _{\beta }$ curves are independent on $\rF$.

Now, consider the regularised shift $E_{2}^{\infty }$, as
introduced in Eq.~(\ref{E12inf}), and the regularised shake-up
diagram $\Lambda _{2\msc{p}}^{\infty }(t)$, reported in
Eq.~(\ref{LambXX}). We need to adjust $\tau _{0}$ for each $\rF$
in such a way that these two quantities are the correct absolute
zero limits for the corresponding numerical quantities, i.e.,
\begin{eqnarray}
\lim_{\beta \hw\rightarrow \infty }
E_{2}^{\beta } &=&E_{2}^{\infty },
\label{E2L2limA} \\
\lim_{\beta \hw\rightarrow \infty }
\Lambda _{2\msc{p}}^{\beta }(t)
&=&\Lambda _{2\msc{p}}^{\infty }(t)\text{.}
\label{E2L2limB}
\end{eqnarray}

\noindent It turns out that $\tau _{0}$ decreases with increasing
$\rF$, which makes sense because the more particles the system has
the more allowed transitions are offered to respond to the sudden
perturbation~(Fig.~\ref{Ffour}\textbf{A}). However, the set of
$\tau _{0}$ values that accurately fit the
condition~(\ref{E2L2limA}) are slightly different from those that
realize the condition~(\ref{E2L2limB}). This is because the
asymptotic forms of $E_{2}^{\infty }$ and $\Lambda
_{2\msc{p}}^{\infty }(t)$ contain terms proportional to
$\rF^{-1/2}$, $\rF^{-3/2}$, $\ldots $, which we have neglected
working in the large $\rF$ limit~(\ref{AppX1}). As a reasonable
compromise, we take the average between these two optimized
sets~(Fig.~\ref{Ffour}\textbf{A}), which is an agreement with both
conditions~(Fig.~\ref{Ffour}\textbf{A}, \textbf{C}, \textbf{D})
within an error less than $5\%$. Interestingly, $E_{2}^{\beta }$
and $E_{1}^{\beta }$ have similar trends and absolute values for
$\alpha =0.2$, while $E_{2}^{\beta }$ is more sensitive to
temperature than $E_{1}^{\beta }$ for $\beta \hw<0.05$. On the
other hand, the modulus of the sub-diagram $\Lambda
_{2\msc{p}}^{\beta }$ presents zeroes at $\w t=k\pi $ and maxima
at $\w t=k\pi /2$, with $k=0,\pm 1,\pm 2,\cdots $. The intensities
of such maxima~(Fig.~\ref{Ffour}\textbf{C}) increase with
increasing the Fermi number~($2\rF$), the critical
exponent~($\alpha $), and the thermal energy~($\beta ^{-1}$). The
phase of $\Lambda _{2\msc{p}}^{\beta}$ is discontinuous at the
extremes of $|\Lambda_{2\msc{p}}^{\beta }|$ and less dependent on
these parameters.

\section{Cumulant expansion}

\label{sec500}

We now have all the ingredients to start our discussion of the
statistical properties of the work distribution of a
non-interacting Fermi gas held in a harmonic trap, following a
sudden local quench of a point like scattering
potential~\cite{sindona:12}. In the present section we will
provide a physical interpretation for the features of the work
distribution. In particular, we will work out the cumulant
expansion of the work distribution and look for the link between
different cumulants and thermodynamical quantities.

\noindent The characteristic function obtained in
Eq.~(\ref{Eq:sudden}) has been introduced in
Eq.~(\ref{eq:loschmidt}) as the mean value of the random variable
$\ee^{i W t/\hbar}$ in the work distribution $\PwQ$ or,
equivalently, as the Fourier transform of the work distribution.
The work distribution, then, is nothing but the absorption
spectrum of the system due to the suddenly switched-on impurity
potential~\cite{sindona:12}: it has been set in Eq.~(\ref{Pwork})
and shown in Fig.~\ref{figTtwo}. We notice that because of the
assumed normalization of $\PwQ$, we have $\chi _{\beta }(0)=1$.

\noindent We call moment of order $n$, or equivalently $n-$th moment of the
distribution, the mean value $\langle W^{n}\rangle $. Once the
characteristic function is known, we can use the differentiation theorem of
Fourier transforms to evaluate each of the above defined moments as%
\begin{equation}
\langle W^{n}\rangle =\left( -{i}\hbar \right) ^{n}\frac{d^{n}\chi _{\beta
}(t)}{dt^{n}}\bigg |_{t=0}.  \label{eq:moments}
\end{equation}%
This relation holds provided that $\chi _{\beta }(t)$ is
continuous and differentiable $n$-times, with all of the
derivatives vanishing at $ t\rightarrow \pm \infty $ (that we will
see not to be always the case).

It will be more convenient for us to work with the cumulant
expansion of $\ln \chi _{\beta }(t)$ instead of computing the moments of $\chi _{\beta }(t)$.
The cumulants are defined analogously to the moments in
Eq.~\ref{eq:moments} with $\ln \chi _{\beta }(t)$ replacing $\chi _{\beta }(t)$. This will have two advantages. First it will
be easier to characterize the distribution since, as we will see,
$ \ln \chi _{\beta }(t)$ makes it possible to calculate important
quantities such as the mean value, variance and skewness
straightforwardly. Second, in our case
$\chi _{\beta }(t)=\Pi _{n}\ee^{\Lambda _{n}^{\beta \ast }(t)}$ so that we will be able to link
the properties of the distribution~(as given by its mean value,
variance and skewness) to the dynamical functions $\Lambda
_{n}^{\beta }(t)$.

\noindent Using the form given by Eq.~(\ref{Eq:sudden}) for the
characteristic function, we express the cumulant generating
function as
\begin{equation}
\ln \chi _{\beta }(t)=\Lambda _{1}^{\beta \ast }(t)+\Lambda _{2}^{\beta \ast
}(t)=\frac{{i}t}{\hbar }(E_{1}^{\beta }+E_{2}^{\beta })-\frac{\delta _{\beta
}^{2}}{2}\w^{2}t^{2}+\Lambda _{2\msc{p}}^{\beta \ast }(t)\text{.}
\label{lnchi}
\end{equation}%
The first three cumulants are, then, given by
\begin{eqnarray}
\kappa _{1}(\beta ) &=&-{i}\hbar \frac{\partial \ln \chi _{\beta }(t)}{%
\partial t}\bigg|_{t=0}=\langle W\rangle   \label{Kappa1} \\
\kappa _{2}(\beta ) &=&-\hbar ^{2}\frac{\partial ^{2}\ln \chi _{\beta }(t)}{%
\partial t^{2}}\bigg|_{t=0}=\langle W^{2}\rangle -\langle W\rangle
^{2}=\sigma ^{2}  \label{Kappa2} \\
\kappa _{3}(\beta ) &=&i\hbar ^{3}\frac{\partial ^{3}\ln \chi _{\beta }(t)}{%
\partial t^{3}}\bigg|_{t=0}=\langle W^{3}\rangle -3\langle W^{2}\rangle
\langle W\rangle +2\langle W\rangle ^{3}=\varkappa \,\sigma ^{3}
\label{Kappa3}
\end{eqnarray}
where we defined the variance $\sigma ^{2}$ and the skewness $\varkappa $.
In our case, and by considering only two sets of diagrams,
see Eq.~(\ref{lnchi}), the above quantities are simply calculated as
\begin{eqnarray}
\kappa _{1}(\beta ) &=&-{i}\hbar \frac{d\Lambda _{1}^{\beta \ast }(t)}{dt}%
\bigg|_{t=0}-{i}\hbar \frac{d\Lambda _{2}^{\beta \ast }(t)}{dt}\bigg|_{t=0}
\nonumber \\
&=&E_{1}^{\beta }+E_{2}^{\beta }-{i}\hbar \frac{d\Lambda _{2\msc{p}}^{\beta
\ast }(t)}{dt}\bigg|_{t=0},  \label{k1} \\
\kappa _{2}(\beta ) &=&-\hbar ^{2}\frac{d^{2}\Lambda _{2}^{\beta \ast }(t)}{%
dt^{2}}\bigg|_{t=0}=\delta _{\beta }^{2}\hbar ^{2}\w^{2}-\hbar ^{2}\frac{%
d^{2}\Lambda _{2\msc{p}}^{\beta \ast }(t)}{dt^{2}}\bigg|_{t=0},  \label{k2}
\\
\kappa _{3}(\beta ) &=&{i}\hbar ^{3}\frac{d^{3}\Lambda _{2}^{\beta \ast }(t)%
}{dt^{3}}\bigg|_{t=0}={i}\hbar ^{3}\frac{d^{3}\Lambda _{2\msc{p}}^{\beta
\ast }(t)}{dt^{3}}\bigg|_{t=0}  \label{k3}
\end{eqnarray}
We see that the two connected diagrams found previously allow us
to calculate the first three cumulants in a straightforward
manner. In particular they are determined by the analytical
properties of the periodic sub-diagram~(\ref{Lambda2E}) at
$t\rightarrow 0$. Now, while $\Lambda _{2\msc{p}}^{\beta }(t)$ is
a continuous and differentiable function, its higher order time
derivatives, i.e., $d^{n}\Lambda _{2\msc{p}}^{\beta }(t)/dt^{n}$
for $n>1$, are ill-defined in the $t\rightarrow 0$-limit. As a
corollary, we have that the higher-order time derivatives of the
characteristic function are not defined at $t=0$, then
Eq.~(\ref{eq:moments}) lacks formal justification for $n\geq 2$.
This is a direct consequence of the point-like modeling of the
impurity potential. A possible work around of the problem will be
proposed in the following paragraphs by introducing suitable
cut-off frequencies on the perturbation matrix
elements~(\ref{Vrrp}). In particular, we shall use the fact that
the regularized characteristic function~(\ref{nustinf}) admits
cumulants of any order to renormalize the second and third ones
given in Eqs.~(\ref{k2}) and~(\ref{k3}). Indeed, the periodic
subdiagram quoted in Eq.~(\ref{nustinf}) has been proved to
accurately reproduce the numerical expression~(\ref{Lambda2E}) in
the absolute zero-limit~(Fig.~\ref{Ffour}). Then, it makes sense
to match the cumulants, as given by Eq.~(\ref{nustinf}), with the
corresponding quantities obtained from the numerical form of $\ln
\chi _{\beta }(t)$, Eq.~(\ref{nust}). Such a condition does not
alter the physics of the work distribution, while it will allow us
to carry out the analysis of the physical behavior of the
skewness, which will turn out to have an expression independent of
this regularization procedure.

\subsection{The mean value}

\label{work}

First let us consider the mean value~(\ref{k1}). Using the expressions for $%
E_{2}^{\beta }$ and $\Lambda _{2}^{\beta }(t)$ given in Eqs.~(\ref{Delta2B})
and~(\ref{Lambda2E}), respectively, we may write%
\[
\kappa _{1}(\beta )=E_{1}^{\beta }+\alpha \eF\sum\limits_{r{\neq }r^{\prime
}=0}^{\infty }\,f_{r}^{+}\,f_{r^{\prime }}^{-}\varphi _{r\rp}+\frac{{i}%
\alpha \eF}{2 \w }\sum\limits_{r{\neq }r^{\prime }=0}^{\infty
}\,f_{r}^{+}\,f_{r^{\prime }}^{-}\frac{d\psi _{r\rp}^{\ast }(t)}{dt}\bigg|%
_{t=0}.
\]%
Then, considering the identity%
\[
\frac{d\psi _{r\rp}^{\ast }(t)}{dt}\bigg|_{t=0}=2{i} \w \varphi _{r\rp}
\]%
we find%
\begin{equation}
\kappa _{1}(\beta )=E_{1}^{\beta }=\langle W\rangle  \label{k1bet}
\end{equation}%
Interestingly\ Eq.~(\ref{k1bet}) states that the mean work is
given by the first-order energy shift of the quenched Fermi gas.
\noindent Such a relation continues to hold at absolute zero with
the regularized
approximations given by Eq.~(\ref{nustinf}), (\ref{E12inf}), and~(\ref%
{gbetapp}). Indeed, the second order time derivative of the
regularized periodic sub-diagram reads%
\[
\frac{d\Lambda _{2}^{\infty \ast }(t)}{dt}\bigg|_{t=0}=\alpha \frac{d}{dt}%
\ln \frac{{\ee}^{2\tau _{0} \w }-1}{{\ee}^{2 \w \tau _{0}-2 \w {i}t}-1}\bigg|%
_{t=0}=-\frac{2 \w i\alpha {\ee}^{2\tau _{0} \w }}{{\ee}^{2\tau _{0} \w }-1}=%
\frac{-iE_{2}^{\infty }}{\hbar }\text{,}
\]%
so that, using Eq.~(\ref{k1}) again, we get: $\kappa _{1}(\infty
)=E_{1}^{\infty }$.

\noindent Although the calculations leading Eq.~(\ref{k1bet}) have
been derived from an approximated expression for the vacuum
persistence amplitude,  we may prove Eq.~(\ref{k1bet}) to be
formally exact. To see this let us consider the
expression~(\ref{IP}) for the characteristic function together
with the definition of the moments in Eq.~(\ref{eq:moments}):%
\[
\kappa _{1}(\beta )=-i\hbar \frac{\partial }{\partial t}\left\langle {T\ee^{%
\frac{{i}}{\hbar }\int_{0}^{t}\dd\tp\tilde{V}(\tp)}}\right\rangle \bigg|%
_{t=0}=\left\langle \tilde{V}(t)T\ee^{\frac{{i}}{\hbar }\int_{0}^{t}\dd\tp%
\tilde{V}(\tp)}\right\rangle \bigg|_{t=0}=\langle \hat{V}\rangle .
\]%
This result tells us that for a sudden quench the average work
done is the mean value of the perturbation after the switch-on at
instant $t=0$.
However, this is also the first-order energy reported in Eq.~(\ref{eq:E1}):%
\begin{eqnarray*}
\langle \hat{V}\rangle &=&\sum_{r,\rp,\xi }V_{r,\rp}\langle c_{2r\,\xi
}^{\dag }c_{2\rp\,\xi }\rangle =\sum_{r,\rp,\xi }V_{r,\rp}\langle c_{2r\,\xi
}^{\dag }c_{2r\,\xi }\rangle \delta _{r\rp} \\
&=&\spind\sum_{r}V_{rr}f_{r}^{+}=E_{1}^{\beta }.
\end{eqnarray*}

\noindent The behavior of $E_{1}^{\beta }$, and hence of the average work $%
\left\langle W\right\rangle $ vs the inverse thermal energy $\beta $, as
been thoroughly discussed in Sec.~\ref{Sec4.2} and shown in Fig.~\ref%
{figTtwo}\textbf{A}

\subsection{The variance}

Using Eq.~(\ref{k2}), we may now evaluate the variance of the work
distribution:%
\[
\sigma ^{2}=\delta _{\beta }^{2}\hbar ^{2} \w ^{2}+\frac{\alpha \hbar \eF}{2 %
\w }\sum\limits_{r{\neq }\rp=0}^{\infty }\,f_{r}^{+}\,f_{\rp}^{-}\frac{%
d^{2}\psi _{r\rp}^{\ast }(t)}{dt^{2}}\bigg|_{t=0}\text{.}
\]%
Here we replace $\delta _{\beta }$ with its explicit form given in Eq.~(\ref%
{eq:gbeta}), and compute the second order time derivative%
\[
\frac{d^{2}\psi _{r\rp}^{\ast }(t)}{dt^{2}}\bigg|_{t=0}=4 \w ^{2}\gamma
_{r}\gamma _{\rp}\text{,}
\]%
which leads the compact expression%
\begin{equation}
\sigma ^{2}=2\alpha \eF\hbar \w \sum\limits_{r,\rp=0}^{\infty }\gamma
_{r}\gamma _{\rp}\,f_{r}^{+}\,f_{\rp}^{-}.  \label{SigSq}
\end{equation}%
To provide an interpretation to this relation, we replace the definition of $%
\alpha $ and the expression~(\ref{Vrrp}) for the impurity potential matrix
elements:%
\[
\sigma ^{2}=\spind\sum\limits_{r=0}^{\infty }f_{r}^{+}\sum\limits_{\rp%
=0}^{\infty }\left\vert V_{r\rp}\right\vert ^{2}f_{\rp}^{-}.
\]%
By analogy with the Fermi's Golden rule, we can look at the quantity%
\begin{equation}
T_{r\rightarrow \rp}=\frac{2\pi }{\hbar }\rho _{0}\left\vert V_{r\rp%
}\right\vert ^{2}f_{\rp}^{-},
\end{equation}%
with $\rho _{0}=2^{-1/2}\hbar ^{-1/2} \w  ^{-1/2}$ denoting the density of
even fermion states in the continuous limit. We recognize $T_{r\rightarrow %
\rp}$ to be the rate of transition from the occupied one-particle state $%
\left\vert 2r\right\rangle $ to the empty state $\left\vert 2\rp%
\right\rangle $. Following this interpretation, we rewrite the variance as%
\begin{equation}
\sigma ^{2}=\frac{\hbar \spind}{2\pi \rho _{0}}\sum\limits_{r}f_{r}^{+}T_{r%
\rightarrow \text{any}}.  \label{Sig0}
\end{equation}%
It is thus clear that $\sigma ^{2}$ gives information on the broadening due
to Fermions scattering in empty states and thus on the spectrum of the
system.

\noindent What is also evident from Eq.~(\ref{SigSq}) is that the sum over
unoccupied particle states, i.e., the $\rp$-series, does not converge,
because the perturbation $V_{r\rp}$ is a weakly decreasing function of $r$
and $\rp$. Indeed, $\sigma ^{2}$ is proportional to the initial product of
the auxiliary functions $\lambda _{\pm }^{\beta }(t)$, introduced in Eq.~(%
\ref{lambdas}):%
\begin{equation}
\sigma ^{2}=2\alpha \eF\hbar \w\,\lambda _{+}^{\beta }(0)\lambda _{-}^{\beta
}(0)\text{.}  \label{SigSqL}
\end{equation}%
Now the asymptotic behaviors $f_{r}^{-}\approx 1$ and $\gamma
_{r}\approx
r^{-1/2}$ for $r\gg 1$ lead to $\lambda _{-}^{\beta }(0)\rightarrow \infty ~$%
(see~\ref{AppX1}). On the other hand, we can substitute
regularized characteristic function~(\ref{nustinf}) in
Eq.~(\ref{k2}) to get the
asymptotic trend%
\begin{equation}
\kappa _{2}(\beta \hw\gg 1)\approx \frac{4\alpha \hbar ^{2}\w^{2}{\ee}%
^{2\tau _{0}\w}}{\left( {\ee}^{2\tau _{0}\w}-1\right) ^{2}}+2\alpha g_{\beta
}\hbar ^{2}\w^{2}.
\end{equation}%
This result, combined with the fact that $g_{\beta }$ vanishes at
the absolute zero, gives
\begin{equation}
\kappa _{2}(\infty )\approx \frac{4\alpha \hbar ^{2}\w^{2}{\ee}^{2\tau _{0}\w%
}}{\left( {\ee}^{2\tau _{0}\w}-1\right) ^{2}}  \label{k2inf}
\end{equation}%
We see that the divergent behavior of $\sigma ^{2}$ in
Eq.~(\ref{SigSqL}) is absorbed by the regularization parameter
$\tau _{0}$. Indeed, considering gases with very large particle
numbers, i.e., working in the $\tau _{0}\w\ll 1$-limit, we find
$\kappa _{2}(\infty )\approx \alpha \hbar ^{2}\tau _{0}^{-2}$.

\noindent To recover a consistent definitions of $\kappa _{2}(\beta )$, as
given by Eq.~(\ref{k2}) and hence Eq.~(\ref{SigSqL}), we proceed similarly
to the Nozi\`{e}res and De Dominicis~\cite{Nozieres} work around of the
Fermi-edge singularity; we introduce an exponential frequency cut-off on the
impurity potential and change the $\gamma _{r}\ $factors to $\gamma _{r}{\ee}%
^{-2r\w/\wo}$. This is equivalent to adding a lifetime width $1/\wo$ to the
unperturbed propagator~(\ref{prop}). By doing so, Eq.~(\ref{SigSq}) becomes
well defined and factorisable as%
\begin{equation}
\sigma ^{2}=2\alpha \eF\hbar \w\,\lambda _{+}^{\beta }(i/\wo)\lambda
_{-}^{\beta }(-i/\wo)\text{,}  \label{SigSqR}
\end{equation}%
and the auxiliary functions are evaluated on the complex time-domain. In~\ref%
{AppX1} we show the cuf-off frequencies $\wo$ is related to the
regularization time $\tau _{0}$ by the condition that $\kappa _{2}(\infty )$%
, as calculated from Eq.~(\ref{SigSqR}) with $\beta \rightarrow
\infty $, matches with the regularized expression~(\ref{k2inf}).

\subsection{The skewness}

The skewness $\varkappa =\kappa _{3}\left( \beta \right) /\kappa _{2}(\beta
)^{3/2}$ is related to both the second an the third cumulant of the
distribution. To begin, we address our attention to the third cumulant~(\ref%
{k3}), i.e.,
\[
\kappa _{3}(\beta )=-\frac{{i}\hbar ^{2}\alpha \eF}{2 \w }\sum\limits_{r{%
\neq }\rp=0}^{\infty }f_{r}^{+}\,f_{r^{\prime }}^{-}\frac{d^{3}\psi
_{rr^{\prime }}^{\ast }(t)}{dt^{3}}\bigg|_{t=0}\text{,}
\]%
keeping in mind that $\kappa _{2}(\beta )$ is a non negative quantity. We
then compute the third order time derivative%
\[
\frac{d^{3}\psi _{rr^{\prime }}^{\ast }(t)}{dt^{3}}\bigg|_{t=0}=-8i \w %
^{3}(r-\rp)\gamma _{r}\gamma _{\rp}
\]%
and write%
\begin{equation}
\kappa _{3}(\beta )=4\alpha \eF\hbar ^{2} \w ^{2}\sum\limits_{r{\neq }\rp%
=0}^{\infty }(r-\rp)\gamma _{r}\gamma _{\rp}\,f_{r}^{+}\,f_{\rp}^{-}.
\end{equation}
Next, we use the definitions for the impurity potential matrix elements and
the transition rates, as in Eq.~(\ref{Sig0}), and express%
\begin{eqnarray}
\kappa _{3}(\beta ) &=&2\spind\hbar \w \sum\limits_{r{\neq }\rp=0}^{\infty
}(r-\rp)\left\vert V_{r\rp}\right\vert ^{2}\,f_{r}^{+}\,f_{\rp}^{-}
\nonumber \\
&=&\frac{\spind\hbar }{2\pi \rho _{0}}\sum\limits_{r}\eF f_{r}^{+}T_{r%
\rightarrow \text{any}}-\frac{\spind\hbar }{2\pi \rho _{0}}\sum\limits_{r}%
\eF f_{r}^{-}T_{\text{any}\rightarrow r},  \label{kk3comp}
\end{eqnarray}
Thus, for the skewness we get
\[
\varkappa \propto\sum\limits_{r}f_{r}^{+}T_{r\rightarrow \text{any}%
}-\sum\limits_{r}f_{r}^{-}T_{\text{any}\rightarrow r},
\]
which encompasses a direct thermodynamical meaning. The first term
is nothing but the energy taken from the system by emptying its
states whereas the second is the energy given to the system by
filling its empty states following the thermodynamic
transformation, which in our case is a sudden quench. So this
quantity tells us whether the transformation effect is to increase
or decrease the internal energy of the gas and is clearly related
to the asymmetry of the work distribution
\begin{figure}[tbh]
\hskip6pc \scalebox{0.99}{\includegraphics{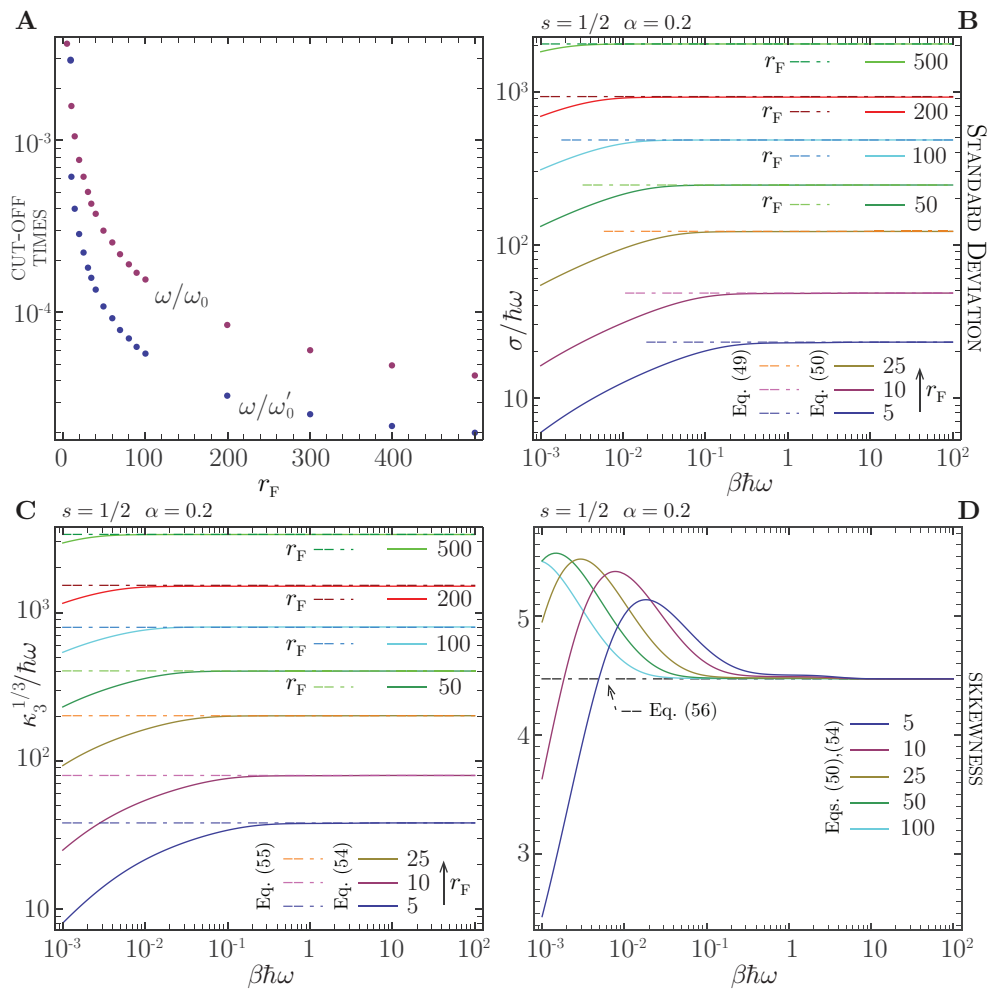}} \vskip-12pt
\caption{(color on line) (\textbf{A}) Adjusted values of the cut-off times $%
1/\wo$~(Purple dots) and $1/\uo$~(Blue dots) vs $\rF$ that let the numerical
expressions~(\ref{k2infapp}) and~(\protect\ref{k3infapp}) for $%
\protect\kappa _{2}(\protect\beta )$ and $\protect\kappa _{3}(\protect\beta %
) $ match with the regularized values given by
Eqs.~(\protect\ref{k2inf})
and~(\protect\ref{k3infreg}), respectively. (\textbf{B}) Variance $\protect%
\sigma =\protect\sqrt{\protect\kappa _{2}(\protect\beta )}$, (\textbf{C})
third cumulant $\protect\kappa _{3}(\protect\beta )^{1/3}$, and (\textbf{D})
skewness $\varkappa $ of the work distribution $\PwQ$ due to a spin-$1/2$
gas, computed with the cut-off times shown in panel~(\textbf{A}). The
curves, plotted vs $\protect\beta \hw$, for $\rF=5-500$ and $\protect\alpha %
=0.2$, tend to the absolute zero approximations obtained
from~(\protect\ref{k2inf}),~(\protect\ref{k3infreg}),
and~(\protect \ref{xiinfreg}).} \label{FigCumul}
\end{figure}

\noindent To provide a consistent definition of $\kappa _{3}(\beta )$, we
need to tackle the divergent behavior of the $\rp$-series in Eq.~(\ref%
{kk3comp}), which is more evident by expressing the third cumulant as%
\begin{equation}
\kappa _{3}(\beta )=2i\alpha \eF\hbar ^{2}\w\frac{d}{dt}\lambda _{+}^{\beta
}(t)\lambda _{-}^{\beta }(t)\bigg|_{t=0}.  \label{kk3lamb}
\end{equation}%
To avoid this divergence, we adopt the previous procedure once
again and redefine Eq.~(\ref{kk3comp}) by adding an imaginary
time-shift to
the auxiliary functions~$\lambda _{\pm }^{\beta }(t)$, i.e.,%
\begin{equation}
\kappa _{3}(\beta )=2i\alpha \eF\hbar ^{2}\w\frac{d}{dt}\lambda _{+}^{\beta
}(t\pm i/\uo)\lambda _{-}^{\beta }(t\pm i/\uo)\bigg|_{t=0}.
\label{kk3cut-off}
\end{equation}

The new cuf-off frequency $\uo$ is fixed by the condition that $\kappa
_{3}(\infty )$, as calculated from Eq.~(\ref{kk3cut-off}) with $\beta
\rightarrow \infty $, tends to the regularized expression%
\begin{equation}
\kappa _{3}(\infty )\approx 8\alpha \,\hbar ^{3}\w^{3}\,e^{2\w\tau _{0}}%
\frac{\left( e^{2\w\tau _{0}}+1\right) }{\left( e^{2\w\tau _{0}}-1\right)
^{3}},  \label{k3infreg}
\end{equation}%
computed with the absolute zero distribution~(\ref{nustinf}).
In~\ref{AppX1} we provide details on how the renormalization of
the second and third
cumulants of the work distribution, given by Eqs.~(\ref{SigSqR}) and~(\ref%
{kk3cut-off}), are carried out. In Fig,~\ref{FigCumul}\textbf{A}
we show the cut-off times $1/\wo$ and $1/\uo$ that let us match
the numerical behavior of $\kappa _{2}(\beta )$ and $\kappa
_{3}(\beta )$ with the expressions given by Eqs.~(\ref{k2inf})
and~(\ref{k3infreg}), respectively. Using such values we can
obtain the behavior of the variance and the
skewness vs $\beta \hbar \w$ for different values of $\rF$~(Fig,~\ref%
{FigCumul}\textbf{B} and~\ref{FigCumul}\textbf{D}). The dependence
of the second and third moment on $\alpha $ is trivial, as both
$\kappa _{2}(\beta ) $ and $\kappa _{3}(\beta )$ are directly
proportional to the critical index. Then, the skewness goes like
$\alpha ^{-1/2}$. Interestingly, we observe that by
Eqs.~(\ref{k2inf}) and~(\ref{k3infreg}) gives the following form
to the skewness parameter%
\begin{equation}
\varkappa \approx \frac{e^{-\w\tau _{0}}}{\sqrt{\alpha }}+\frac{e^{\w\tau
_{0}}}{\sqrt{\alpha }}\text{,}  \label{xiinfreg}
\end{equation}%
which does not suffer from the divergent behavior of $\kappa _{2}(\infty )$
and $\kappa _{3}(\infty )$. Furthermore, in the $\tau _{0}\w\ll 1$-limit,
the third cumulant behaves as $\kappa _{3}(\infty )\approx 2\alpha \hbar
^{3}\tau _{0}^{-3}$ and the skewness turns out to depend only on the
critical exponent $\varkappa \approx 2/\sqrt{\alpha }$~(Fig.~\ref{FigCumul}%
\textbf{D}). Being completely independent of the regularization
procedure, this is a physically meaningful result, showing how the
critical parameter $\alpha$ does indeed determine the asymmetry of
the work distribution, ultimately due to the Fermi edge behavior.

\section{Irreversible work}

\label{sec600}

In the previous section we have seen that the work distribution
contains information about both the unperturbed system and the
system following the thermodynamic transformation. In this section
we shall first recover a Jarzynski-like equality, which will help
us in identifying the hypothetical final equilibrium state of the
system to compare with. We shall then find an approximate and
analytic expression, and discuss a numerical method for the
computation of the irreversible work, which is a figure of merit
of the irreversibility of the transformation. We are interested in
highlighting the microscopic origin of irreversibility for the
sudden quench studied above. First let us recover the
Jarzynski-like equality.

\noindent We start from the evaluation of $\left\langle \ee^{-\beta
W}\right\rangle $ by Eq.~(\ref{eq:qworkdist}), i.e.,%
\begin{eqnarray}
\left\langle \ee^{-\beta W}\right\rangle  &=&\frac{1}{Z}\sum\limits_{n,m}\ee%
^{-\beta (E_{n}+\mu N_{n})}\,p_{m|n}\,\ee^{-\beta (E_{m}^{\prime }-E_{n})}=%
\frac{1}{Z}\sum\limits_{n,m}\ee^{\mu \beta N_{n}}\,p_{m|n}\,\ee^{-\beta
E_{m}^{\prime }}  \nonumber \\
&=&\frac{1}{Z}\sum\limits_{m}\ee^{\mu \beta N_{m}^{\prime }}\,\ee^{-\beta
E_{m}^{\prime }}  \label{eeW1}
\end{eqnarray}%
In going from the first to the second row we have used particle
number conservation, $p_{m|n}\propto \delta _{N_{n},N_{m}^{\prime
}}$, and the completeness relation $\sum_{n}p_{m|n}=1$,
\cite{Tasaki}. We recognize that the $n$-sum in Eq.~(\ref{eeW1})
is nothing but the grand canonical partition function for the gas
in equilibrium with the impurity potential, with same temperature
and chemical potential as the initial one. Hence, we have:%
\begin{equation}
\left\langle \ee^{-\beta W}\right\rangle =\frac{Z^{\prime }}{Z}\equiv \ee%
^{-\beta (\Omega ^{\prime }-\Omega )},  \label{eeWW}
\end{equation}%
where
\begin{equation}
\Omega =-\frac{1}{\beta }\ln \,Z=-\frac{\spind}{\beta }\sum\limits_{n=0}^{%
\infty }\ln \left[ 1+\ee^{-\beta \left( \varepsilon _{n}-\mu \right) }\right]
\label{Omega0}
\end{equation}%
and
\begin{equation}
\Omega ^{\prime }=-\frac{1}{\beta }\ln \,Z^{\prime }=-\frac{\spind}{\beta }%
\sum\limits_{n=0}^{\infty }\ln \left[ 1+\ee^{-\beta \left( \varepsilon
_{n}^{\prime }-\mu \right) }\right]   \label{OmegaPP}
\end{equation}%
denote the grand potentials for the unperturbed and perturbed
equilibrium states, respectively. The fact that the Jarzynski
relation makes explicit connection with a (hypothetical)~final
equilibrium state is meaningful as it gives us a reference for the
(hypothetical)~reversible version of the transformation
encompassed by the switching on of the external perturbation. The
reversible version of this protocol, thus, would involve a change
in the number of particles, which is necessary in order to
maintain the initial chemical potential even if the single
particle energies are modified during the protocol \cite{gici}.

By means of the Jensen inequality $\left\langle \ee^{-\beta W}\right\rangle
\geq \ee^{-\beta \left\langle W\right\rangle }$ we can derive the statement
of the second law of thermodynamics :%
\begin{equation}
\left\langle W\right\rangle -\left( \Omega ^{\prime }-\Omega \right) \geq 0
\label{JensIneq}
\end{equation}%
As expected, the average work done on the system is greater than
the change in the grand potential of the initial state and the
hypothetical final equilibrium state. This relation suggests us to
define a new variable that we shall name $W_{\msc{irr}}=W-\Delta
\Omega $, with $\Delta \Omega =\Omega ^{\prime }-\Omega $. The new
variable will have the very same distribution as the original one
but for the mean value which will be shifted by an amount $\Delta
\Omega $:
\begin{equation}
\left\langle W_{\msc{irr}}\right\rangle =\left\langle W\right\rangle -\Delta
\Omega ,  \label{Wirr}
\end{equation}

In the following, we propose an analytic and a numerical
approaches to compute this excess work~(\ref{Wirr}) and discuss it
in order to highlight the nature of irreversibility in our system.
We have already set
all the elements to compute $\Omega $, i.e., the unperturbed level energies $%
\er$, distribution functions $f_{r}^{\pm }$, and chemical
potential $\mu $, and we have already discussed the behavior of
the average work $\langle W\rangle $~(see
Fig.~\ref{figTtwo}\textbf{A}). What is left to calculate is the
perturbed grand potential $\Omega ^{\prime }$, for which we will
provide both an approximate analytic and a numerical computation.

\subsection{Perturbation approach}

The energies entering the hypothetical final state belong to the spectrum of
the Hamiltonian%
\begin{equation}
H(x)=H_{0}(x)+\pi V_{0}x_{0}\delta (x),  \label{HT}
\end{equation}%
which describes a particle of the gas in equilibrium with both the
harmonic trap and the impurity potential. As discussed in
Sec.~\ref{sec200}, the odd single particle energy levels are left
unperturbed $\varepsilon _{2r+1}^{\prime }=\varepsilon _{2r+1}$.
As for the even energies $\varepsilon _{2r}$, which cannot be
calculated analytically, we resort to perturbation theory assuming
the height $\pi V_{0}x_{0}$ of the $\delta $-potential to be small
with respect to the unperturbed energies. Then, we express:
\begin{equation}
\varepsilon _{2r}^{\prime }\approx \varepsilon _{2r}+\delta \varepsilon
_{2r},  \label{E2rPert}
\end{equation}%
where $\delta \varepsilon _{2r}=\sum_{i=1}^{\infty }\varepsilon _{2r}^{(i)}$
is computed from the non-degenerate Rayleigh-Schr\"{o}dinger perturbation
theory%
\begin{equation}
\varepsilon _{2r}^{(1)}=V_{0}\gamma _{r},\,\varepsilon
_{2r}^{(2)}=V_{0}^{2}\gamma _{r}\sum\limits_{r^{\prime }\neq r}\frac{\gamma
_{r^{\prime }}}{\varepsilon _{2r}-\varepsilon _{2r^{\prime }}},\ldots
\label{RSPertTh}
\end{equation}%
based on the assumption%
\begin{equation}
\varepsilon _{2r}\gg \varepsilon _{2r}^{(1)}\gg \varepsilon _{2r}^{(2)}\gg
\ldots   \label{assumpt}
\end{equation}%
Substituting~(\ref{E2rPert}) in the perturbed grand potential~(\ref%
{OmegaPP}), and carrying out a power series expansion for small $\delta
\varepsilon _{2r}$, to the second order we get:%
\[
\Omega ^{\prime }=\Omega +(2s+1)\sum\limits_{r=0}^{\infty }f_{r}^{+}\delta
\varepsilon _{2r}-\frac{\beta (2s+1)}{2}\sum\limits_{r=0}^{\infty
}f_{r}^{+}f_{r}^{-}\left[ \delta \varepsilon _{2r}\right] ^{2}
\]%
Now, considering the assumption~(\ref{assumpt}), we find%
\begin{eqnarray}
\Delta \Omega  &\approx &(2s+1)\sum\limits_{r=0}^{\infty
}f_{r}^{+}\varepsilon _{2r}^{(1)}+(2s+1)\sum\limits_{r=0}^{\infty
}f_{r}^{+}\varepsilon _{2r}^{(2)}  \nonumber \\
&&-\frac{\beta (2s+1)}{2}\sum\limits_{r=0}^{\infty }f_{r}^{+}f_{r}^{-}\,%
\left[ \varepsilon _{2r}^{(1)}\right] ^{2}.
\end{eqnarray}%
By Eq.~(\ref{RSPertTh}), the first term at the right hand side of this
relation is simply the average work $\left\langle W\right\rangle $
calculated above~(see Sec.~\ref{work}). The second term is the second order
contribution $\Delta U^{(2)}$ to the change in the total energy of the
system. The third one may be expressed in terms of the Gaussian broadening~(%
\ref{eq:gbeta}) as $\beta \delta _{\beta }^{2}\hbar ^{2}\omega ^{2}/2$. We
are now able to calculate the excess work~(\ref{Wirr}) as
\begin{equation}
\left\langle W_{\msc{irr}}\right\rangle \approx \frac{\delta _{\beta }^{2}}{2%
}\beta \hbar ^{2}\omega ^{2}-\Delta U^{(2)}.  \label{WirrPert}
\end{equation}%
Notice that by Eq.~(\ref{RSPertTh}) the second-order corrections $%
\varepsilon _{2r}^{(2)}$ are negative, which makes $-\Delta
U^{(2)}$ a positive quantity, and let the Jensen
inequality~(\ref{JensIneq}) be always verified.
Eq.~(\ref{WirrPert}) can be brought into a more interesting and
general form, by noticing that%
\begin{eqnarray*}
\sum\limits_{r,\xi }\left[ \varepsilon _{2r}^{(1)}\right]
^{2}f_{r}^{+}f_{r}^{-} &=&\sum\limits_{r,r^{\prime },\xi ,\xi ^{\prime
}}\varepsilon _{2r}^{(1)}\varepsilon _{2r^{\prime }}^{(1)}\left(
f_{r}^{+}f_{r^{\prime }}^{+}+f_{r}^{+}f_{r}^{-}\delta _{rr^{\prime }}\delta
_{\xi \xi ^{\prime }}\right)  \\
&=&\sum\limits_{r,r^{\prime },\xi ,\xi ^{\prime }}\varepsilon
_{2r}^{(1)}\varepsilon _{2r^{\prime }}^{(1)}\left[ \langle \hat{c}_{2r\,\xi
}^{\dag }\hat{c}_{2r\,\xi }\hat{c}_{2r^{\prime }\,\xi ^{\prime }}^{\dag }%
\hat{c}_{2r^{\prime }\,\xi ^{\prime }}\rangle -f_{r}^{+}f_{r^{\prime }}^{+}%
\right]  \\
&=&\mathrm{Var}\left( \sum\limits_{r,\xi }\varepsilon _{2r}^{(1)}\hat{c}%
_{2r\,\xi }^{\dag }\hat{c}_{2r\,\xi }\right)
\end{eqnarray*}%
We thus conclude that the irreversible work thus takes the suggestive form:%
\begin{equation}
\left\langle W_{\msc{irr}}\right\rangle \approx \frac{\beta }{2}\text{Var}%
\left(\sum\limits_{r,\xi }\varepsilon _{2r}^{(1)}\hat{c}_{2r\,\xi }^{\dag }\hat{c}%
_{2r\,\xi }\right)-\Delta U^{(2)}\text{.}  \label{WirrAPP}
\end{equation}%
This relation is independent on the model used to characterize the
impurity potential. Interestingly, we may look at the occupation
numbers $n_{2r\,\xi }=0,1 $ as independent random variables,
distributed according to the probability
distribution $\frac{e^{-\beta (\varepsilon _{2r}-\mu )n_{2r\,\xi }}}{%
1+e^{-\beta (\varepsilon _{2r}-\mu )}}$, the function
\begin{equation}
W[\{n_{r\,\xi }\}]=\sum\limits_{r,\xi }\varepsilon _{2r}^{(1)}n_{2r\,\xi }
\label{W1def}
\end{equation}%
of the configurations $\{n_{r\,\xi }\}$ is a random variable too.
As a rsult, the more peaked this random function is the smaller
its contribution to the irreversible work. This means that for an
adiabatic change there is no spread of the work distribution,
since in that limit the average work done would be a Dirac delta
function of $\Delta \Omega $.

\subsection{Numerical Approach}

We now turn the attention to the numerical calculation of\ the perturbed
grand potential~(\ref{OmegaPP}), which requires the knowledge of the
eigenvalues of the single-particle Hamiltonian~(\ref{HT}). The odd harmonic
oscillator eigenfunctions $\psi _{2r+1}(x)$ and eigenenergies $\varepsilon
_{2r+1}$ are left unaffected by the $\delta $-potential, due to the fact
that $\psi _{2r+1}(0)=0$~(see Sec.~\ref{sec200}). On the other hand the
perturbed even eigenfunctions of~(\ref{HT}), with the physically correct
asymptotic behavior, are the parabolic cylinder functions
\begin{equation}
\psi _{2\kr}(x)=\frac{\eta _{\kr}\,D_{2\kr}\left( \sqrt{2}|x|/x_{0}\right) }{%
\pi ^{1/4}x_{0}^{1/2}\Gamma (2\kr+1)^{1/2}},
\end{equation}%
with associated level energies $\ek=\hw(2\kr+1/2)$. The latter are the just
perturbed energies denoted $\varepsilon _{2r}^{\prime }$ above. As shown for
example in Ref.~\cite{Goold:2008}, the stationary Shr\"{o}dinger equation for
the $\delta $-potential implies
\[
\frac{d\psi _{2\kr}}{dx}\bigg|_{x\rightarrow 0^{+}}-\frac{d\psi _{2\kr}}{dx}%
\bigg|_{x\rightarrow 0^{-}}=\frac{2\pi V_{0}}{x_{0}\hw}\psi _{2\kr}(0),
\]%
where
\[
\frac{d\psi _{2\kr}}{dx}\bigg|_{x\rightarrow 0^{\pm }}=\mp \frac{2\Gamma
\left( 1/2-\kr\right) }{x_{0}\Gamma (-\kr)}\psi _{2\kr}(0).
\]%
\begin{figure}[tbh]
\hskip6pc \scalebox{0.99}{\includegraphics{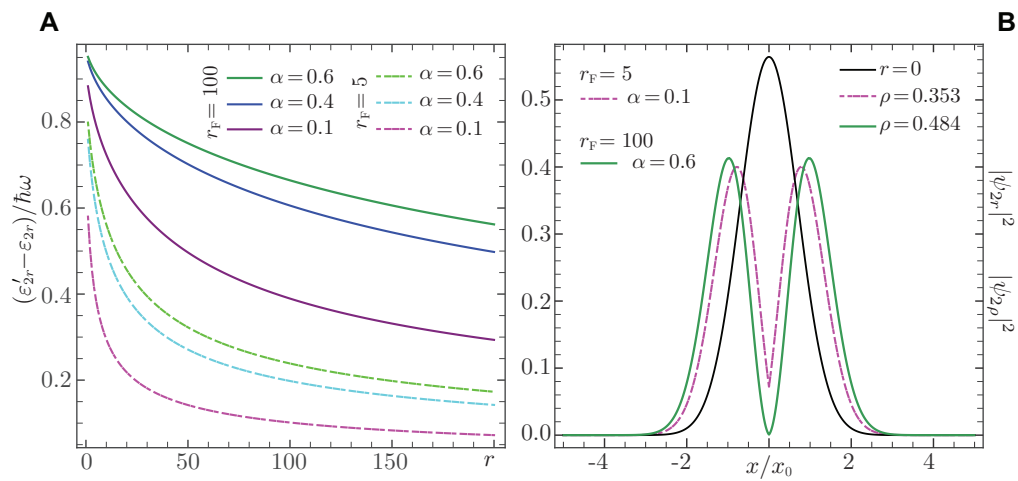}} \vskip-12pt
\caption{(Color online) Single particle energy shifts $\ek-\er$ in units of $
\hw$ (\textbf{A}) and Spatial probability densities $|\protect\psi _{2r}|^{2}
$ and $|\protect\psi _{2\kr}|^{2}$ associated to the single-particle ground
state (\textbf{B}). The $\protect\delta $-potential strength is calculated
from the values $\protect\alpha =0.1,0.4,0.6$, $\rF=5,100$, which
corresponds to different values of $\kr$ in (\textbf{B})}
\label{wk}
\end{figure}
The last two relations yield an implicit condition between the strength of
the $\delta $-function and the quantum numbers $\kr$:
\begin{equation}
-\frac{\pi V_{0}}{2\hw}=\frac{\Gamma (1/2-\kr)}{\Gamma (-\kr)}.
\label{Implicit}
\end{equation}%
Since the $\Gamma $-function has poles for negative integer values, Eq.~(\ref%
{Implicit}) leads to $\kr\rightarrow r$ for $V_{0}\rightarrow 0$, and $\kr%
\rightarrow r+1/2$ for $V_{0}\rightarrow \infty $. Then, the
energy eigenvalues $\ek$ converge to the unperturbed energies
$\varepsilon _{2r}$ when the potential barrier is set to zero,
while they tend to $\varepsilon _{2r+1}$ for an infinite barrier.
In the latter case, the perturbed energies become identical to the
values of the odd eigenfunctions, leading to a double degeneracy
of all eigenvalues. For arbitrary values of $V_{0}$, which means
for $\alpha = 0-1$ at any $\rF>0$, see Eq.~(\ref{alphaPP}), we
need to solve Eq.~(\ref{Implicit}) numerically and get a sequence of
quantum numbers $\kr=\kr(\alpha ,\rF)$, with $r{\leq }\kr\leq
r+1/2$. Here, it is important then to notice that for each fixed
value of $\alpha $ and $\rF$ there exists a one to one
correspondence between each unperturbed quantum number $r$ and a
$\kr$. Once the $\kr$ are known, we can compute the perturbed
energies $\ek$ and the normalization constants $\eta _{\kr}=\eta
(\alpha ,\rF)$, which let us
determine the perturbed wavefunctions $\psi _{2\kr}(x)$. In Fig.~\ref{wk}%
\textbf{A} and~\ref{wk}\textbf{B}, we show how the single particle ground
state wave functions and the energy levels change with increasing both $%
\alpha $ and $\rF$.
\begin{figure}[tbh]
\hskip6pc \scalebox{0.99}{\includegraphics{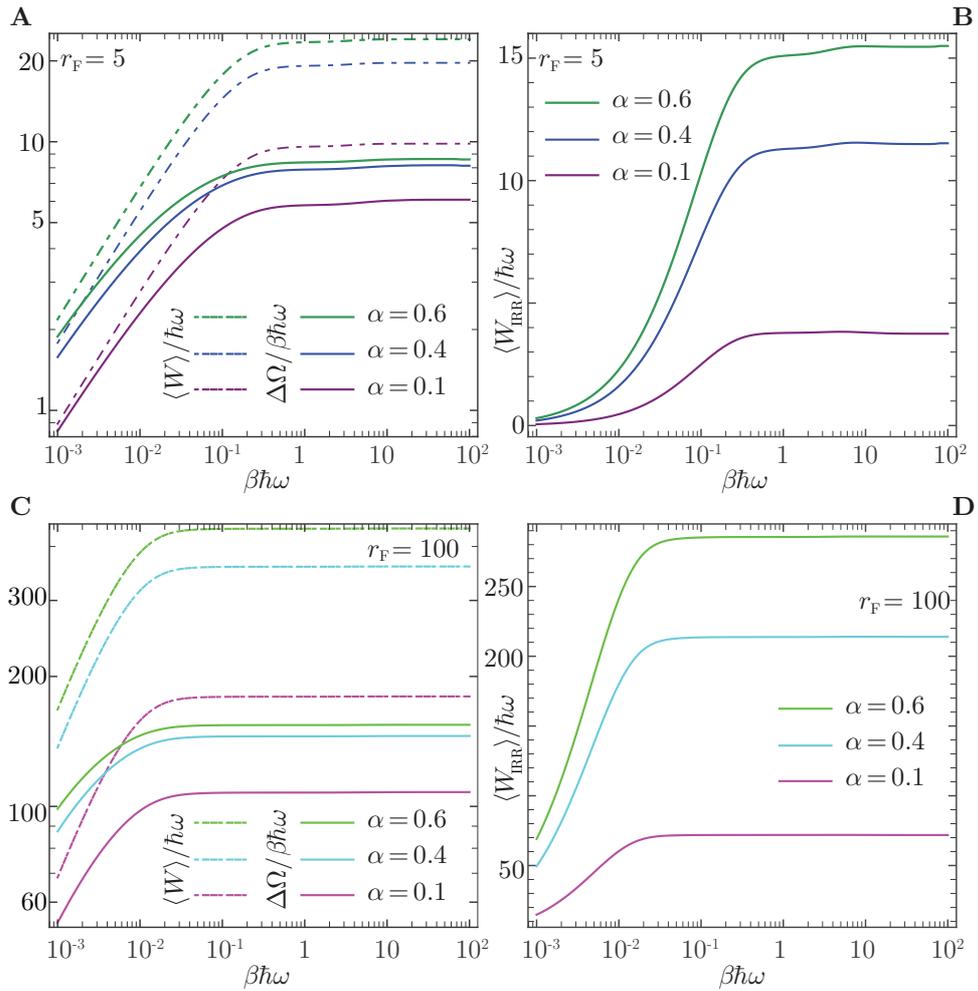}} \vskip-12pt
\caption{(Color online) $\protect\beta \hbar \protect\omega $-dependence of
the Average work, the grand potential variation and the irreversible work,
as computed from Eqs.~(\protect\ref{eq:E1}), (\protect\ref{k1bet}), (\protect
\ref{NumIrr}), and~(\protect\ref{Wirr}). The critical parameter is allowed
to take the values $\protect\alpha =0.1$, $0.4$, $0.6$, in a spin-$1/2$ gas
low and large particle numbers, i.e., $\rF=5,100$. In panels~\textbf{A} and~%
\textbf{C} the Jarzynski inequality~(\protect\ref{JensIneq}) is proved to
hold for each sampled values of $\protect\alpha $ and $\rF$. In panels~%
\textbf{B} and~\textbf{D} the excess of work done by the system is shown to
reach a value independent on the temperature of the system for $\protect%
\beta \hw\gtrsim 0.01-0.1$, depending on $\rF$. Such a value increases with
both $\rF$ and $\protect\alpha $.}
\label{WoWo}
\end{figure}

Within this framework, the perturbed grand-canonical potential is given by%
\begin{eqnarray*}
\Omega ^{\prime } &=&-\frac{\spind}{\beta }\sum\limits_{r=0}^{\infty }\ln %
\left[ 1+\ee^{-\beta \left( \varepsilon _{2r+1}-\mu \right) }\right]  \\
&&-\frac{\spind}{\beta }\sum\limits_{r=0}^{\infty }\ln \left[ 1+\ee^{-\beta
\left( \ek-\mu \right) }\right].
\end{eqnarray*}%
The irreversible work can be thus computed from~(\ref{Wirr})
using%
\begin{equation}
\beta \Delta \Omega =\spind\sum\limits_{r=0}^{\infty }\ln \left[ \frac{1+\ee%
^{-\beta \left( \varepsilon _{2r}-\mu \right) }}{1+\ee^{-\beta \left( \ek%
-\mu \right) }}\right] .  \label{NumIrr}
\end{equation}%
In Fig.~\ref{WoWo}\textbf{A} and~\ref{WoWo}\textbf{C} we see that the
second law~(\ref{JensIneq}) is obeyed for any chosen valued
of $\alpha $ and $\rF$. The amount of irreversible work, shown in Fig.~\ref%
{WoWo}\textbf{B} and~\ref{WoWo}\textbf{D}, increases with
decreasing temperature reaching a nearly constant saturation value
for low enough temperatures, $\beta \hw\gtrsim 0.01-0.1$ depending
on the number of particles in the gas. A similar trend is followed
by other crucial quantities discussed here, such as the chemical
potential, the energy shifts, and the cumulants of the work
distribution
function. On the other hand, as suggested by the approximation~(\ref%
{WirrPert}), the irreversible work is highly sensitive to the total
particle number and the critical exponent following the proportionality
relation $\left\langle W_{\msc{irr}}\right\rangle \propto \alpha \rF$, for $%
\rF\gg 1$.

\section{Conclusions}
\label{concluse}
In this work, we explored the physics and
thermodynamics of an inhomogeneous Fermi gas perturbed by the
sudden switch on of a local scattering potential. Exploiting the
direct relationship between the characteristic function of work
and the vacuum persistence amplitude, and by means of the linked
cluster expansion technique, we obtained the full statistics of
work done by performing such a local quench on the gas, showing
that the first, second and third moments of the work distribution
encapsulate the salient thermodynamic features of the model. {\bf
Indeed, features of the textbook Fermi-edge singularity problem
were found to be present in the higher moments. Furthermore, we
obtained analytic and perturbative expressions for the excess or
irreversible work done on the system as well as a Jarzynski like
equality.} From a general perspective, our work demonstrates the
potential of combining ideas coming from recent developed
techniques in non-equilibrium statistical mechanics with
traditional approaches from many body physics for the analysis of
out of equilibrium problems in quantum systems. It is also worth
pointing out that the approach developed here is by no means
restricted to a specific system and can be extended to other types
of quench problems, global and local, in non-interacting and
interacting many-particle systems.
\\
In addition to the theoretical framework developed, one can
imagine that in the future the work distribution of the Fermi gas
maybe extracted in an experimental setting by means of coupling to
an auxiliary ancilla system which would then function as a probe,
giving access to the relevant thermodynamic quantities by
monitoring, e.g. its decoherence dynamics \cite{osid}. This was
first suggested in the context of Fermi gases by Goold {\ et al}
in \cite{goold:11}. In fact very recent proposals to verify the
quantum fluctuation relations by means of interferometry of an
ancillary probe qubit \cite{dorner2, mauro} have been realised in
a recent experiment in a NMR setting \cite{nmr}. The challenge for
the future is to really push these experiments to the many-body
domain.

\begin{appendix}
\section{Low thermal energy expansion of the characteristic function of work and its cumulants} \label{AppX1}
In the following we will present the details of the analytical approximations for the quantities introduced in Sec.~\ref{Sec4.1}, i.e., the energy shifts~(\ref{eq:E1}) and~(\ref{Delta2B}), the Gaussian standard deviation~(\ref{eq:gbeta}), and the shake-up sub-diagram~(\ref{Lambda2E}), which determine the characteristic function of work~(\ref{nustinf}). By the same methods, we will show how to find the asymptotic limits for the main properties of the work distribution, i.e.,  its mean value, variance and skewness~(see Sec.~\ref{sec500}).

\noindent To begin, we consider the power series expansions of the statistical Fermi factors
\begin{eqnarray}
f_{r}^{\pm } &=&\sum\limits_{m=0}^{\infty }(-1)^{m}\ee^{\pm 2 \w  \tau_{m}(r-r_{\mu })}, \qquad \quad r\lessgtr r_{\mu }, \label{frexp} \\
&=&-\sum\limits_{m=1}^{\infty }(-1)^{m}\ee^{\mp 2 \w  \tau_{m}(r-r_{\mu})}, \qquad r\gtrless r_{\mu }, \nonumber
\end{eqnarray}
as well as their products
\begin{equation}
f_{r}^{+}f_{r}^{-}=\sum\limits_{m=1}^{\infty }(-1)^{m+1}\,m\, \ee^{\pm 2 \w  \tau_{m}(r-r_{\mu })},\qquad r\lessgtr r_{\mu }, \label{f+f-exp}
\end{equation}
with $\tau_{m}$ denoting the characteristic times $\tau_{m}=m\beta \hbar $ induced by thermal fluctuations.

\subsection{Thermal series and low temperature approximation for the one- and two-vertex loops}

By the relations~(\ref{frexp}) and~(\ref{f+f-exp}) we can convert Eqs.~(\ref{eq:E1}), (\ref{Delta2B}), (\ref{eq:gbeta}), and (\ref{Lambda2E}) into power series of $\ee^{\pm 2\beta \hw(r-r_{\mu })}$.
In particular, the auxiliary functions $\lambda_{\pm }^{\beta }(t)$, entering the connected graphs~(\ref{Lambda1}) and~(\ref{Lambda2}), are expanded as
\begin{equation}
\lambda_{\pm }^{\beta }(t)=\sum_{m=0}^{\infty }(-1)^{m}\,\lambda_{m\pm}^{\beta }(t). \label{lambdasexp}
\end{equation}
Using Eq.~(\ref{frexp}), the coefficients of the series~(\ref{lambdasexp}) read:
\begin{eqnarray}
\lambda_{0+}^{\beta }(t) &=&\sum_{r<r_{\mu }}\gamma_{r} \ee^{2 {i} r \w  t}, \quad \lambda_{0-}^{\beta }(t)=\sum_{r>r_{\mu }}\gamma_{r} \ee^{-2 {i} r \w  t} \label{CoeL1} \\
\lambda_{m\pm }^{\beta }(t) &=&\pm \sum_{r<r_{\mu }}\gamma_{r}\ee^{2 \w  \tau_{m}\left( r-r_{\mu }\right) }\ee^{\pm 2 {i} r  \w  t } \mp \sum_{r>r_{\mu }}\gamma_{r}\ee^{-2 \w  \tau_{m}\left( r-r_{\mu }\right) } \ee^{\pm 2 {i} r  \w  t }. \label{CoeL2}
\end{eqnarray}
%\subsection{One- and Two-vertex connected diagrams}
From the expansion~(\ref{CoeL2}) we obtain the thermal series for the first-order energy shift, i.e., the average work:
\begin{equation}
\left\langle W\right\rangle =E_{1}^{\beta }=\sqrt{2\spind\hw\eF\alpha} \sum_{m=0}^{\infty }(-1)^{m}\lambda_{m+}^{\beta }(0). \label{Wexp}
\end{equation}
In addition, we rewrite the two-vertex loop~(\ref{Lambda2}) as
\begin{equation}
\Lambda_{2}^{\beta }(t)=-\frac{2\alpha  \w  \eF}{\hbar }\sum_{m,m^{\prime}=0}^{\infty }(-1)^{m+m^{\prime }}\int_{0}^{t}\dd\tp\,\int_{0}^{\tp}\dd\ts\,\lambda_{m+}^{\beta }(\ts){\,}\lambda_{m^{\prime }-}^{\beta }(\ts), \label{L2ExP}
\end{equation}
which by Eq.~(\ref{L2decomp}) contains all other basic quantities of the problem, see Eqs.~(\ref{Delta2B}), (\ref{eq:gbeta}), and (\ref{Lambda2E}).
To compute the Gaussian term, however, it is more straightforward to work on expression given in Eq.~(\ref{eq:gbeta}) and use the expansion~(\ref{f+f-exp}).
By doing so, we get
\[\Lambda_{2\msc{g}}^{\beta }(t)=-\alpha g_{\beta } \w  ^{2}t^{2},\qquad g_{\beta }=\sum_{m=1}^{\infty }(-1)^{m+1}\,m\,g_{m}^{\beta }, \]
in which
\[ g_{m}^{\beta }=\frac{\eF}{\hw}\sum_{r<r_{\mu }}\gamma_{r}^{2}\ee^{-2 \w \tau_{m}(r_{\mu }-r)}+\frac{\eF}{\hw}\sum_{r>r_{\mu }}^{\infty }\gamma_{r}^{2}\ee^{-2 \w  \tau_{m}(r-r_{\mu })}. \]
As a first approximation, we focus on the temperature range where the chemical potential is approximated by Eq.~(\ref{muinf}), i.e., $\beta \hw\gtrsim 0.2$ for $\rF>5$.
In this regime, we apply the sum rules
\begin{eqnarray}
\sum_{r=r_{1}}^{r_{2}}\gamma_{r}\,z^{r} &=&z^{r_{1}}\,\HypFR(r_{1},z)-z^{r_{2}+1}\,\HypFR(r_{2}+1,z),\qquad z\neq 1, \nonumber \\
&=&\left( 2r_{2}+1\right) \,\gamma_{r_{2}}-2r_{1}\,\gamma_{r_{1}},\qquad
\qquad \qquad \quad z=1, \label{FinSums}
\end{eqnarray}
with $\HypFR$ being the regularised Hypergeometric function
\begin{equation}
\HypFR(r,z)=\HypF(1,1/2+r,1+r;z)\,\gamma_{r}=\sum_{m=0}^{\infty }\gamma_{m+r}\,z^{m}. \label{HypINF}
\end{equation}
Substituting into Eqs.~(\ref{CoeL1}) and~(\ref{CoeL2}), we find
\begin{eqnarray}
\lambda_{0+}^{\beta }(t) &\approx &\frac{\sqrt{\pi }}{\sqrt{1-\ee^{2 {i}  \w  t }}}-\ee^{2 {i}  \w  t \left( \rF+1\right) }\HypFR\left( \rF+1,\ee^{2 {i}  \w  t }\right),\label{Coel1INF}\\
\lambda_{0-}^{\beta }(t) &\approx &\ee^{2 {i}  \w  t \left( \rF +1\right) }\HypFR\left( \rF+1,\ee^{-2 {i} \w  t }\right),\label{Coel2INF} \\
\lambda_{m\pm }^{\beta }(t) &\approx &\pm \frac{\sqrt{\pi }\ee^{-\beta m\eF}}{\sqrt{1-\ee^{2 \w  \left( \pm {i} t+\tau_{m}\right) }}} \label{LambdHyp}
\\
&&\mp \ee^{\pm 2 {i}  \w  t \left( \rF+1\right) }\bigg\{\ee^{3\tau_{m} \w  /2}\HypFR\left[ \rF+1,\ee^{2 \w  \left( \pm {i} t+\tau_{m}\right) }\right] \nonumber \\
&&\qquad \qquad \qquad \qquad +\ee^{-3\tau_{m} \w  /2}\HypFR\left[ \rF+1, \ee^{2 \w  \left( \pm {i} t-\tau_{m}\right) }\right] \bigg\} \nonumber
\end{eqnarray}
As a second approximation, we take systems with large numbers of particles, and employ the large-$\rF$ expansions:
\begin{equation}
\gamma_{\rF\gg 1}=\rF^{-1/2}+\mathrm{o}(\rF^{-3/2}),\qquad \HypFR(\rF\gg1,z)=\frac{\rF^{-1/2}}{1-z}+\mathrm{o}(\rF^{-3/2})\text{,} \label{SpecFAS}
\end{equation}
so that we straightforwardly get
\begin{eqnarray}
\lambda_{0+}^{\beta }(t) &\approx &\frac{\sqrt{\pi }}{\sqrt{1-\ee^{2 {i}  \w  t}}}-\frac{\ee^{2 {i}  \w  t \rF}}{\sqrt{\rF}\left( 1-\ee^{2 {i}  \w  t }\right) },\\ \lambda_{0-}^{\beta }(t)&\approx &\frac{\ee^{-2 {i}  \w  t \rF}}{\sqrt{\rF}\left( 1-\ee^{-2 {i}  \w  t }\right)}. \label{CoeL1AS} \\
\lambda_{m\pm }^{\beta }(t)&\approx & \mp \frac{\ee^{2 {i}  \w  t \left(\rF+1\right) }}{\sqrt{\rF+1}}\left[ \frac{\ee^{3\tau_{m} \w  /2}}{1-\ee^{2 \w  \left( \pm {i} t+\tau_{m}\right) }}+\frac{\ee^{-3\tau_{m} \w  /2}}{1-\ee^{2 \w  \left( \pm {i} t-\tau_{m}\right) }}
\right]. \label{CoeL2AS}
\end{eqnarray}
Now, the first-order energy shift~(\ref{Wexp}) is approximated by Eqs.~(\ref{CoeL1AS}) and~(\ref{CoeL2AS}), as:
\begin{equation*}
E_{1}^{\beta }\approx 2\sqrt{\alpha \spind}\eF+\frac{\hw\sqrt{\alpha \spind}}{\ee^{3\beta \hw/2}+1}\left( 2\ee^{\frac{\beta \hw}{2}}+\frac{3\ee^{\frac{3\beta \hw}{2}}}{2}+\frac{3}{2}-2\ee^{\beta \hw}\right).
\label{E1betaExpINF}
\end{equation*}
This expression is largely dominated by the $\beta $-independent value, reported in Eq.~(\ref{E12inf}) and shown in Fig.~\ref{FigThree}\textbf{A}.
Turning to the Gaussian coefficient, a simple change of summation indices, with $\mu =\mu_{\infty }$, leads to
\[
g_{m}^{\beta }=
\frac{\eF}{\hw}
\ee^{- \w  \tau_{m}/2}\sum_{r=0}^{\rF}\gamma_{\rF-r}^{2}
\ee^{-2 \w  \tau_{m}r}+
\frac{\eF}{\hw}
\ee^{ \w  \tau_{m}/2}
\sum_{r=1}^{\infty }\gamma_{\rF+r}^{2}\ee^{-2 \w  \tau_{m}r}.
\]
The transformed summations in this last line are dominated by low $r$ terms.
In a many fermion environment, by the asymptotic relation~(\ref{SpecFAS}), we approximate $\gamma_{\rF\pm r}^{2}\approx \gamma_{\rF}^{2}\approx \rF^{-1} $ and neglect terms going like $\ee^{-2 \w  \tau_{m}\rF}$, to obtain
\[
g_{m}^{\beta }\approx 2\frac{\ee^{ \w  \tau_{m}/2}+\ee^{3 \w  \tau_{m}/2}-\ee^{-2 \w  \tau_{m}\rF}}{\ee^{2 \w  \tau_{m}}-1}\approx 2\frac{\ee^{ \w  \tau_{m}/2}}{\ee^{ \w  \tau_{m}}-1},
\]
so that
\[
g_{\beta }\approx 2\sum_{m=1}^{\infty }(-1)^{m}m\frac{\ee^{ \w  \tau_{m}/2}}{\ee^{ \w  \tau_{m}}-1}.
\]
This relation, being identical to Eq.~(\ref{gbetapp}), allows us
to express express
\begin{equation}
\delta_{\beta }\approx 2\alpha ^{1/2}\left[ \sum_{m=1}^{\infty }(-1)^{m}m\frac{\ee^{ \w  \tau_{m}/2}}{\ee^{ \w  \tau_{m}}-1}\right] ^{1/2}.
\label{deltabAPP}
\end{equation}
Then, the standard deviation at low temperatures is independent
of the number of particles in the
gas~(Fig.~\ref{FigThree}\textbf{B}).

\noindent As for the second-order shift~$\Lambda_{2\msc{s}}^{\beta
}(t)$ and the Fermi-edge component~$\Lambda_{2\msc{p}}^{\beta
}(t)$, we first consider the $\beta $-independent
coefficients~(\ref{CoeL1AS}) of the series~(\ref{lambdasexp}),
write down the product
\begin{equation}
\lambda_{+}^{\infty }(t)\lambda_{-}^{\infty }(t)\approx \frac{\sqrt{\pi }\ee^{-2 {i} \rF  \w  t}}{\sqrt{\rF}\left( 1-\ee^{-2 {i}  \w  t}\right) ^{3/2}}-\frac{1}{\rF\left( 1-\ee^{-2 {i}  \w  t}\right) ^{2}},
\label{lambdasexpP}
\end{equation}
and plug it into the time-ordered integrals~(\ref{L2ExP}). Then,
we need to deal with the short-time singularity of
Eq.~(\ref{lambdasexpP}) by adding an imaginary time regularization
to the $\ts$-integral, i.e., we need to shift the $\ts$
integration domain by ${i} \tau_{0}$. The resulting integral is
dominated by the Fermi-edge terms reported in Eq.~(\ref{LambXX}):
\begin{eqnarray}
\Lambda_{2}^{\infty }(t) &=& {i}  \w  t\frac{2\alpha \ee^{2\tau_{0} \w  }}{\ee^{2\tau_{0} \w  }-1}+\alpha \ln \left( \frac{\ee^{2\tau_{0} \w  }-1}{\ee^{2 \w  (\tau_{0}+ {i} t)}-1}\right) +\mathrm{o}\left( \rF^{-1/2}\right) \text{.} \label{LambFES}
\end{eqnarray}
\begin{figure}[b]
%\raggedleft
\hskip 6pc \scalebox{0.99}{\includegraphics{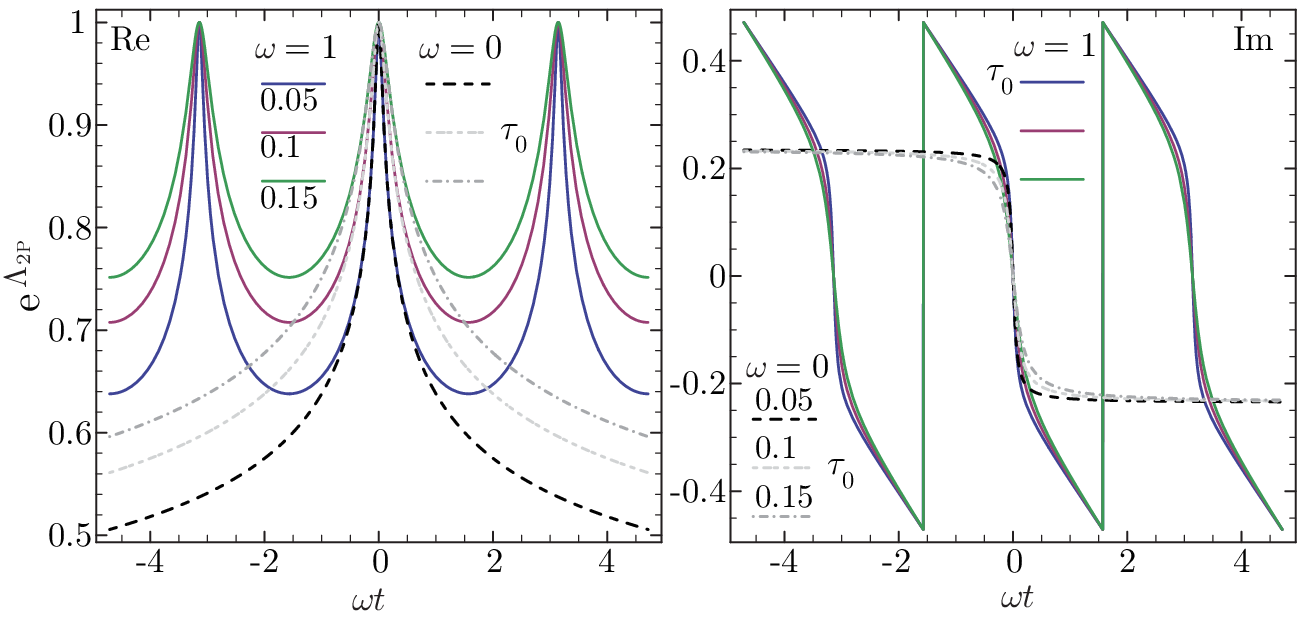}} \vskip -12pt
\caption{Real and imaginary parts of the excited impurity propagator $\ee^{\Lambda_{2\msc{p}}^{\infty }(t)}$ in the trapped~($ \w =1$) and free~($ \w  \to 0$) fermion gas. The critical exponent is fixed to $\alpha=0.1$, while several regularisation times~($\tau_0=0.05, 0.1, 0.15$) are tested.}
\label{FAPP1}
\end{figure}
Since at the absolute zero the Gaussian part is absent, we may readily interpret
\begin{equation}
\Lambda_{2\msc{s}}^{\infty }(t)\approx {i}  \w  t\frac{2\alpha \ee^{2\tau_{0} \w  }}{\ee^{2\tau_{0} \w  }-1}\quad \mathrm{and} \quad \Lambda_{2\msc{p}}^{\beta }(t)\approx \alpha \ln \left( \frac{\ee^{2\tau_{0} \w  }-1}{\ee^{2 \w  (\tau_{0}+ {i} t)}-1}\right) .
\label{Lambda0TT}
\end{equation}
As shown in Figs.~\ref{Ffour}\textbf{B}-\textbf{D}, these results
are in excellent agreement with the numerical calculations at
$\beta \hw\rightarrow \infty $ by suitable adjustments of
$\tau_{0}$. Such a regularization parameter depends on $\rF$ and
takes the physical interpretation of the average time needed by
the system to respond to the abrupt impurity perturbation at zero
temperature~(Fig.~\ref{Ffour}\textbf{A}). It is also interesting
to note that if we let the harmonic frequency in
Eq.~(\ref{Lambda0TT}) go to zero, by fixing $\alpha $ and keeping
the number of particles in the gas~($2\rF \approx \eF/\hw$)
finite, we retrieve the Nozieres-De Dominicis
result~\cite{mahan:00,Nozieres,sindona:12}
\begin{equation}
\Lambda_{\msc{mnd}}(t)=-\alpha \ln (it/\tau _{0}+1),
\label{MNDLamb}
\end{equation}
which leads to the propagator $\ee^{\Lambda_{\msc{mnd}}(t)}=\left(
it/\tau _{0}+1\right)^{-\alpha}$, originally calculated for a
suddenly switched on core-hole in a free electron gas. Such a
limiting procedure is  illustrated in Fig.~\ref{FAPP1}, where we
clearly see that as $ \w  \to 0$, the periodicity of the
propagator $\ee^{\Lambda_{2\msc{p}}^{\infty }(t)}$ vanishes and
its modulus reduces to a single peak with tails going like $1/t$.
Inclusion of finite temperature corrections~\cite{sindona:12} may
be done by considering all possible products of $\lambda_{0\pm
}^{\infty }(t)$ and $\lambda_{m>0\pm }^{\infty }(t)$, as given by
Eq.~(\ref{LambdHyp}), then using the expansion~(\ref{HypINF}), and
finally performing the $\tp$ and $\ts$-integrals excluding terms
proportional to $t^{2}$. However, the zero temperature
result~(\ref{Lambda0TT}) is largely dominant in characteristic
function $\chi_{\beta} (t)$, and in the system response
$\nu_{\beta }(t)$, within the ranges of temperatures~$\beta
\hw=0.4-\infty $ and particle numbers~$\langle \hat{N}\rangle
\gtrsim 20$ considered here.
\subsection{Characteristic function and its cumulants at the absolute zero}
We now use the approximations~(\ref{Wexp}), (\ref{deltabAPP}), and
(\ref{Lambda0TT}) in Eq.~(\ref{nust}), so that the characteristic
function of the work  distribution reads%
\begin{eqnarray}
\chi _{\beta }(t) &=&{\ee}^{\frac{{i}t}{\hbar }(E_{1}^{\beta }+E_{2}^{\beta
})}{\ee}^{-\frac{\delta _{\beta }^{2}}{2}\w^{2}t^{2}}{\ee}^{\Lambda _{2%
\msc{p}}^{\beta \ast }(t)}  \label{chiINF1} \\
&\approx &{\ee}^{2i\sqrt{\alpha \spind}\frac{\eF t}{\hbar }}{\ee}^{\frac{%
-2i\alpha \w t}{1-\ee^{-2\tau _{0}\w}}}  \nonumber \\
&&\qquad \times {\ee}^{-2\alpha \w^{2}t^{2}\sum_{m}(-1)^{m}m%
\frac{\ee^{\w\tau _{m}/2}}{\ee^{\w\tau _{m}}-1}}\left( \frac{{\ee}^{2\tau
_{0}\w}-1}{{\ee}^{2\w\tau _{0}-2\w {i}t}-1}\right) ^{\alpha },
\label{chiINF2}
\end{eqnarray}%
where the last line is equivalent to Eq.~(\ref{nustinf}) of the main text.
In Sec.~\ref{Sec4.2}, we have observed that Eq.~(\ref{chiINF2}) is indeed a
low temperature limit for the numerical expression~(\ref{chiINF1}) once the
regularisation parameter $\tau _{0}$ is adjusted in order to obey the
conditions~(\ref{E2L2limA}) and~(\ref{E2L2limB}). The adjusted values of $%
\tau _{0}$ have been also reported in Fig.~\ref{Ffour}\textbf{A}\ vs $\rF$.

In Sec.~\ref{sec500} we have provided a method to determine the first three
cumulants of the work distributions, see Eqs.~(\ref{Kappa1})-(\ref{Kappa3}).
In particular, we have seen that these quantities depend on the auxiliary
functions~$\lambda _{\pm }^{\beta }(t)$ and their first-order
time-derivatives at the time the impurity potential is activated. The first
cumulant is well defined, being proportional to $\lambda _{+}^{\beta }(0)$,
and coincides with the first-order energy shift~(\ref{Wexp}) discussed
above, as reported in Eq.~(\ref{k1bet}). On the other hand the second and
third cumulants involve $\lambda _{-}^{\beta }(0)$ and $d\lambda _{-}^{\beta
}(t)/dt|_{t=0}$, see Eqs.~(\ref{SigSqL}) and~(\ref{kk3lamb}). These two
quantities contain weighted sums of the $\gamma _{r}$ and $r\gamma _{r}$%
-factors by the hole occupation numbers~$f_{r}^{-}$. For large
$r$, the two series diverge like $r^{-1/2}$ and $r^{1/2}$,
respectively. It is not surprising that a normalizable
distribution function, with a well defined mean value, has
divergent higher order moments, and this is due to both the sudden
dynamics and the spatial modelling of the impurity
potential. However, the regularization procedure introduced in Sec.~\ref%
{Sec4.2}, and applied above to derive Eq.~(\ref{chiINF2}), leads
to a well defined characteristic function admitting finite
cumulants of any order. To have a consistent theory, we also want
the cumulants of the work distribution, as computed from
Eq.~(\ref{chiINF1}), to correctly tend to the same values obtained
from Eq.~(\ref{chiINF2}). With this in mind, we have proposed in
Sec.~\ref{sec500} a renormalization procedure in which the
auxiliary functions are extended on the complex-time domain
$t\rightarrow t+i\tau $ and shifted by a small imaginary time,
i.e., $\tau =\pm 1/\w_{0}$
in Eq.~(\ref{SigSqR}) and $\tau =\pm 1/\w_{0}^{\prime }$ in Eq.~(\ref%
{kk3cut-off}). Then, using Eqs.~(\ref{CoeL1}), (\ref{Coel1INF}) and~(\ref%
{Coel2INF}) above, we can use the extended auxiliary functions $\lambda
_{\pm }^{\infty }(t+i\tau )$ to obtain the absolute zero expressions
\begin{eqnarray}
\kappa _{2}(\infty ) &=&2\alpha \eF\hbar \w\,\lambda _{+}^{\infty
}(0)\lambda _{-}^{\infty }(0)  \nonumber \\
&=&2\alpha \eF\frac{\sqrt{\pi }e^{-2\w(\rF+1)/\wo}}{\sqrt{1-e^{-2\w/\wo}}}\,%
\HypFR(\rF+1,e^{-2\w/\wo})  \nonumber \\
&&\qquad \qquad -2\alpha \eF e^{-4\w\left( \rF+1\right) /\wo}\,\HypFR(\rF%
+1,e^{-2\w/\wo})^{2}  \label{k2infapp}
\end{eqnarray}%
and
\begin{eqnarray}
\kappa _{3}(\infty ) &=&2i\alpha \eF\hbar ^{2}\w\frac{d}{dt}\lambda
_{+}^{\beta }(t)\lambda _{-}^{\beta }(t)\bigg|_{t=0}  \nonumber \\
&=&\frac{2\alpha \eF e^{-2\w(\rF+1)/\w_{0}^{\prime }}}{\sqrt{\pi }\sqrt{%
1-e^{-2\w/\uo}}}\left[ 2(\rF+1)+\frac{1}{1-e^{-2\w/\uo}}\right] \HypFR(\rF%
+1,e^{-2\w/\uo})  \nonumber \\
&&+\frac{4\sqrt{\pi }\alpha \eF e^{2\w(\rF+2)/\uo}}{\sqrt{1-e^{-2\w/\uo}}}%
\frac{d}{dz}\HypFR(\rF+1,z)\bigg|_{z=e^{-2\w/\uo}}.  \label{k3infapp}
\end{eqnarray}%
To link the these expressions with the regularized ones that we
have derived in Sec.~\ref{sec500}, we have fixed the cut-off
frequencies $\wo$ and $\uo$ by constraining:
\begin{equation*}
\text{Eq.~(\ref{k2infapp})}=\text{Eq.~(\ref{k2inf})}
\qquad \text{and} \qquad
\text{Eq.~(\ref{k3infapp})}=\text{Eq.~(\ref{k3infreg})}
\end{equation*}
The adjusted values of the corresponding cut-off times $1/\wo$ and $1/\uo$ have been reported in Fig.~\ref{FigCumul}\textbf{A}.
\end{appendix}

\end{document}